\documentclass[twocolumn]{aastex631}
\usepackage{hyperref}
\usepackage{booktabs}
\usepackage{graphicx}
\usepackage{amsmath}
\usepackage{array}

\accepted{to AJ}
\begin{document}

\title{Demography of stellar radio population within 500~pc: A VLASS-\textit{Gaia DR3} study}

\correspondingauthor{Ayanabha De}
\email{astronomerrik@gmail.com}

\correspondingauthor{Mayank Narang}
\email{mayankn1154@gmail.com}

\author[0000-0001-8845-184X]{Ayanabha De}
\affiliation{Department of Astronomy \& Astrophysics, Tata Institute of Fundamental Research,
Homi Bhabha Road, Colaba,
Mumbai, 400005, India}

\author[0000-0002-0554-1151]{Mayank Narang}
\affiliation{Academia Sinica Institute of Astronomy \& Astrophysics, 11F of Astro-Math Bldg., No.1, Sec. 4, Roosevelt Rd., Taipei 10617, Taiwan, R.O.C.}

\author[0000-0002-3530-304X]{P.Manoj}
\affiliation{Department of Astronomy \& Astrophysics, Tata Institute of Fundamental Research,
Homi Bhabha Road, Colaba,
Mumbai, 400005, India}

\author[0000-0002-2585-0111]{B. Shridharan}
\affiliation{Department of Astronomy \& Astrophysics, Tata Institute of Fundamental Research,
Homi Bhabha Road, Colaba,
Mumbai, 400005, India}

\author[0000-0002-9497-8856]{H.Tyagi}
\affiliation{Department of Astronomy \& Astrophysics, Tata Institute of Fundamental Research,
Homi Bhabha Road, Colaba,
Mumbai, 400005, India}

\author[0000-0001-8075-3819]{Bihan Banerjee}
\affiliation{Department of Astronomy \& Astrophysics, Tata Institute of Fundamental Research,
Homi Bhabha Road, Colaba,
Mumbai, 400005, India}

\author[0000-0002-4638-1035]{Prasanta K. Nayak}
\affiliation{Instituto de Astrofísica, Pontificia Universidad Católica de Chile, Av. Vicuña MacKenna 4860, 7820436, Santiago, Chile}

\author[0000-0002-9967-0391]{Arun Surya}
\affiliation{Indian Institute of Astrophysics, 2nd block Koramangala, Bangalore, 560034, India}



\begin{abstract}

In this work, we have carried out a systematic analysis of the VLASS quick look catalogs together with \textit{Gaia DR3} to identify the optical counterparts of 3~GHz radio emitters within 500~pc to obtain a homogeneous statistical sample of stellar radio sources. We have identified distinct populations of 3 GHz emitters across the \textit{Gaia DR3} color-magnitude diagram. We also present candidate sources (transient, highly variable or background artifacts) which can be confirmed by follow-up observations. A majority of the detected sources constitute main sequence G, K and M-type stars including ultra-cool dwarfs. Pinning down the origin of radio emission from these populations can help us gain further insights into the origin of stellar and planetary magnetic fields. By analyzing the variation of brightness temperature of the sources with their spectral type, we have tentatively associated possible emission mechanisms with different object types. We inspected the correlation between quiescent radio and X-ray emission for our sample that can provide crucial insights into the current understanding of the Gudel-Benz relationship, which is essential for modeling steady radio emission and coronal heating. This VLASS-\textit{\textit{Gaia DR3}} analysis acts as a pilot study for follow-up observations at multiple wavelengths to better understand stellar structure, model flaring activities and detect radio emission caused by star-planet interactions.
\end{abstract}


\section{Introduction} \label{sec:intro}

Radio emission from stars and sub-stellar objects offers a unique window into their physical characteristics, magnetic fields, and interaction with their surroundings. Observations across different wavelengths along with modeling the emission can contribute significantly to our understanding of these objects. By analyzing near-simultaneous observations at multiple frequencies across several epochs at radio wavelengths,  plasma and magnetic field properties in the stellar surroundings and emission mechanisms can be studied \citep{gudel_review, dulk}. Radio observations can probe the stellar chromospheres, coronae, winds, and accretion around stars and YSOs \citep{bookbinder,Das_2022,Vedantham_emission}. Radio observation is also the only unambiguous method for detecting exoplanet magnetic fields \citep{shkolnik+08,Cauley+19,Narang_2020}. 

The radio brightness temperature (defined in Section \ref{Tbright}) along with effective temperature measurements of the source and spectral index estimates, can help derive the nature of emission and source optical depth \citep{gudel_review}. A study of how the emission properties vary across different spectral and object types can shed light on the physics of stellar structure and evolution. 

Stellar radio emission results from various physical processes associated with different objects and spectral types. Non-thermal radio emission in main sequence low mass ($0.5M_{\odot}\leq M_* \leq 1.5M_{\odot}$, i.e late F to early M type) stars, is mostly driven by persistent magnetic activity generated due to the presence of tachocline (boundary between radiative and convective layers) and differential rotation in outer convective layer  \citep{dwarfs1}. The presence of magnetic field in late M dwarfs and brown dwarfs is not well understood since they lack tachocline  \citep{dwarfs2}. Large-scale magnetospheric dynamics are likely the origin of radio emission from M dwarfs and brown dwarfs  \citep{vlass_new_hv}. Hot B, A and early F type stars ($ M_* \geq 1.5M_{\odot}$) with almost entirely radiative interiors have no intrinsic magnetic fields  \citep{star_mag}. However, there are magnetic chemically peculiar stars (MCPs) that might have retained their fossil magnetic fields. Radio emission from such stars is possibly wind driven  \citep{wind}.

Close-in exoplanets can form Jupiter-Io like systems with their host stars  \citep{Kavanagh_spi}, thus providing a method to measure exoplanetary magnetic fields  \citep{spi}. Search for such emissions is an active field in radio astronomy  (e.g.,\cite{Etangs+11,Narang_2020,GJ1151,2021RNAAS...5..158N,yzceti,2024MNRAS.529.1161N,vol_lim_exo24}). Similar to Jupiter-Io coupling  \citep{Jupiter-Io}, an exomoon can trigger activity in the magnetosphere of its host exoplanet generating low-frequency radio emission \citep{Exomoon0,exomoon,2023MNRAS.522.1662N}.  

To study the origin and properties of stellar radio emission from various object types, several targeted observations and surveys have been carried out. For example, \cite{RSCVn} carried out a survey to observe radio bursts from RS Canum Venaticorum (RS CVn) and Algol binaries.  Similarly \cite{MCP} and \cite{Das_2022} carried out extensive observations of MCPs to detect radio emissions from them.  \citet{WTTS} detected coherent bursts from weak-line T-Tauri stars (WTTS). Emission from the corona of the M-dwarf WX Uma was studied in a target-specific observation by  \cite{wxuma}. Volume-limited radio surveys have been carried out for OB type radio stars (e.g., \cite{OBsurvey}) and for ultra-cool dwarfs (e.g.,\cite{UCD, Berger_2002}). 

Targeted radio observations can be biased towards known radio-bright sources. To study the statistical properties of different radio populations, volume-limited unbiased wide field sky surveys are necessary. Wide-field surveys like the Faint Images of the Radio Sky at Twenty-one centimeters (FIRST) \citep{first}, the NRAO VLA Sky Survey (NVSS) \citep{nvss}, the TIFR GMRT Sky Survey (TGSS) \citep{tgss} and the Westerbork Northern Sky Survey (WENSS) \citep{wenss} have detected and characterized many new radio objects \citep{Kimball_2008, first_sdss}. However, the low angular resolution and sensitivity and large astrometric uncertainties of these sky surveys increase the probability of chance alignment with background galaxies and false detection of artifacts  (e.g.,\cite{first_sdss}) and require extensive follow-up (e.g.,\cite{2022MNRAS.515.2015N}). The Very Large Array Sky Survey (VLASS) \citep{vlass}, the Rapid ASKAP Continuum Survey (RACS) \citep{racs}, the LoFAR Two Meter Sky Survey (LoTSS)  \citep{lotss} and the Galactic and Extra-galactic all sky MWA survey (GLEAM) \citep{gleam} are some of the recent radio surveys with higher sensitivity, astrometric accuracy and increased resolution which should reduce chance alignment probabilities and false detections.

Using the catalogs published from the aforementioned surveys, recent works have identified new radio stars \citep{pm_method} and compiled new catalogs for megahertz to gigahertz stellar radio sources \citep{Sydney,vlass_new_hv}. Most of the sources identified through their analysis are low-frequency emitters. VLASS provides a window into the decimetric radio sky, where we expect many magnetic stars to emit gyrosynchrotron radio waves and young stellar objects to be present \citep{gudel_review}. \cite{vlass_new_hv} conducted a statistical analysis of certain stellar populations and their properties using VLASS, LoTSS, and the Gaia Catalog of Nearby Stars (GCNS) \citep{GCNS}. However, their analysis focused on studying the variation of detection rates with spectral types, so they limited their sample to a 50 pc volume in which GCNS is complete for all objects earlier than M8. This sample primarily consists of M-dwarfs, for which they derived flare statistics.

In this work, we aim to study the general properties of broad range of radio emitters - ranging from ultra-cool dwarfs to hot B-types, from binary systems to young stellar objects. For this investigation we use the first two epoch data of VLASS along with \textit{Gaia DR3} to identify radio population within 500~pc. Such a large sample based on homogeneous selection criteria covering diverse spectral types, allows us to study the nature of emission and their variation across different spectral and object types. Two epochs of observation separated by 32 months with the same sensitivity also allow variability studies. Our study focuses on identifying diverse populations of radio emitting stellar systems in decimetric wavelength bands and characterizing the origin of their emission. We discuss stellar radio emission in the context of stellar structure and environment. In Section \ref{sec:samsel} we describe the data sets used in the paper and filtering methods. Our analysis and results are presented in Section \ref{results}, and we discuss our results in Section \ref{discussion}. Section \ref{summary} provides a summary of the paper. 

\section{Data \& Sample selection} \label{sec:samsel}

VLASS is an National Radio Astronomy Observatory (NRAO) initiative to carry out a continuum survey of the entire sky above a declination of $-40^\circ$, using the Karl G. Jansky Very Large Array in B and BnA configuration at S-band (2-4GHz) \citep{vlass}. The project, which was initiated in 2017 and is scheduled for completion by 2024, encompasses three distinct epochs of observation, each separated by an approximate interval of 32 months. The survey has RMS noise of $120\,\mu$Jy per epoch and an estimated $60~\mu$Jy RMS noise for the 3 epoch stacked images with an angular resolution of $\sim$ $2.5\arcsec{}$. \textit{Gaia} is an ESA all sky survey mission  \citep{gaia_dr3} that provides most precise astrometric and spectro-photometric measurements for $\sim 1.7$ billion stars in optical band.

In this work, we have cross-matched \textit{Gaia $3^{rd}$ Data Release (DR3) catalog} \citep{gaia_dr3} and \textit{VLASS Epoch 1 \& 2 Quick Look (QL1 and QL2) catalogs} \citep{vlassql1} to identify the optical counterparts of the 3 GHz radio sources within 500~pc using precise parallax measurement from \textit{Gaia DR3}. The QL catalogs, produced from minimally cleaned images, are well-suited for demographic studies of radio stellar populations. More deeply cleaned and self-calibrated Single Epoch (SE) images also provide in-band spectral index information, but due to their current incompleteness across the entire VLASS footprint, they are not used in this study. The combination of \textit{Gaia DR3} and VLASS QL catalogs provides us with a large enough volume-limited homogeneous sample to study the demographics of emission properties of the stellar radio population across all spectral and object types. To reduce the possibilities of false associations of radio emitters with background galaxies detected by Gaia, we only consider sources with $parallax\_over\_error > 10$. This ensures that we have removed sources with bad parallax measurements.

\begin{figure*}
  \centering
  \includegraphics[width=0.45\textwidth]{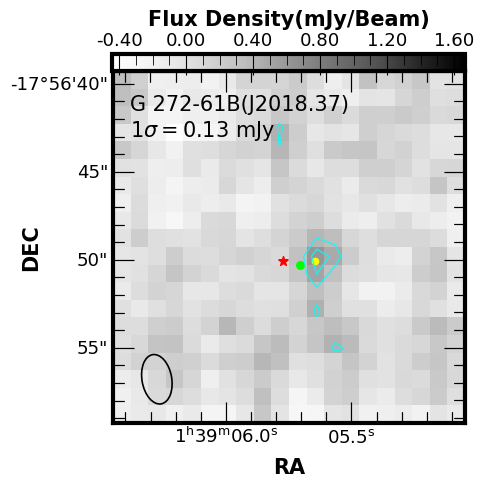}
  \includegraphics[width=0.45\textwidth]{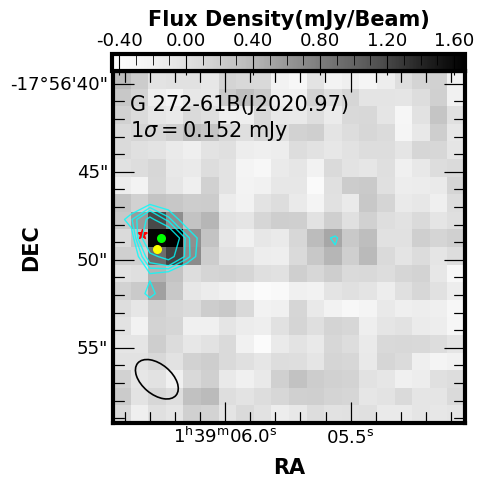}
  \caption{Offset between proper motion propagated \textit{Gaia DR3} position for the eruptive variable G 272-61B, and VLASS component in Epoch 1 and 2 is shown in (a) and (b), respectively. Red star symbol and lime solid circle represent the \textit{Gaia DR3} position of G 272-61B corrected to corresponding VLASS mean epochs and exact observed epochs respectively. The yellow solid circle show the VLASS 1.2 and 2.2 cataloged positions in the respective images. Contours are drawn at 3, 4, 5 and $7\sigma$ levels, $\sigma$ being the local rms noise. The synthesized beam is shown as the black ellipse.}
  \label{wxuma}
\end{figure*}

The first epoch of VLASS observations was carried out from September 2017 to July 2019  and the second epoch was carried out from April 2020 to April 2022. Thus mean epoch for VLASS QL1 is 2018.7 and for VLASS QL2, it is 2021.33. The positions of \textit{Gaia DR3} sources, whose epoch is J2016, were proper motion corrected to J2018.7 and J2021.33. In the catalog user guide\footnote{\href{http://cutouts.cirada.ca/}{of the VLASS Quick Look and Single Epoch Catalogs web-page}}, the positions of VLASS and \textit{Gaia DR2} \citep{gaiadr2} sources after epoch correction, were compared to derive the typical astrometric offsets in VLASS catalogs to be $\sim 0.5\arcsec{}$ above $-20^{\circ}$ declination and up to $1 \arcsec{}$ between $-20^{\circ}$ to $-40^{\circ}$ declinition. Therefore we consider a search radius of $1\arcsec{}$ to account for the VLASS astrometric offset and epoch uncertainties due to proper motions up to $0.5 \arcsec{}/year$. Limiting ourselves to $1\arcsec{}$ search radius reduces the number of background radio components chance-aligning with a \textit{Gaia DR3} source. Using this strategy, however, we can miss sources with proper motion $ \gtrsim 0.5\arcsec{}/year$ (see Section \ref{hpm}).  

Radio galaxies and artifacts present in VLASS catalogs can align by chance with \textit{Gaia DR3} objects within the limited search radius. Artifacts can be filtered using data flags (explained in Section \ref{filter}). To bypass chance alignments with uncatalogued galaxies, several methods can be employed. Identification by filtering polarized sources \citep{pritchard,calingham} is biased to polarized radio sources. Serendipitous searches \citep{meerkat1,meerkat2} are biased towards highly variable sources. Proper motion searches have no bias towards properties of emission \citep{pm_method}. However, the volume limit of proper motion search relies heavily on time baselines between two epochs of observations, and positional uncertainty \citep{2022MNRAS.515.2015N} of the survey, making it biased towards high proper motion stars or limited to small volume searches. To obtain a statistical sample of radio stars across many spectral types, we adopted a simpler version of the proper motion search outlined in \citet{pm_method}; we will call our method of searching as ``Radio source identification by multi-epoch association".

If a radio component is found within the search radius of the position (epoch corrected) of an optical source, they are considered to be potentially associated with each other. If such an association can be found in more than one epoch, the association is considered unambiguous. This strategy captures the high proper motion sources. For low proper motion stars, we are relying on the fact that background artifacts are less likely to randomly chance align with proper motion-corrected positions of optical sources at two or more epochs. Unknown steady galactic sources could still contaminate the final sample with this strategy, and this is a limitation of the results of our cross-match strategy for the low proper motion stars. 

To summarise, We have cross-matched \textit{Gaia DR3} catalog with VLASS QL Catalog 1 and 2 (different epochs) to obtain two samples which we term as \textit{Sample A} and \textit{Sample B} here onwards. If a VLASS component is found within $1\arcsec{}$ of the \textit{Gaia DR3} source, they are considered, tentatively, to be associated with each other. If such a match can be found in both Samples A and B (two epochs), the association is considered robust with little or no ambiguity. Otherwise, the VLASS component could be a background artifact \citep{2022MNRAS.515.2015N} or a highly variable candidate radio source \citep{pm_method}.


\subsection{High Proper-Motion sources}\label{hpm}

The general cross-match strategy discussed above will fail to identify high proper motion radio sources ($>0.5\arcsec{}/yr$) observed towards the beginning or end of a single epoch survey. This is because VLASS takes roughly 2 years to map its entire footprint once and in the general cross-match strategy, we only use the mean epoch. A good example is an eruptive variable G 272-61B as shown in Figure \ref{wxuma}, which has a high proper motion of $3.18 \arcsec{}/year$. It was observed by VLASS in April 2018, therefore its epoch is $2018.37$. After proper motion correction to J2018.7 (mean of QL1), the \textit{Gaia DR3} position is still off by $\sim 1.5 \arcsec{}$. With a search radius (tolerance limit) of $1\arcsec{}$, the true optical counterpart for this radio source couldn't have been identified. Therefore, we have separately dealt with the high proper motion \textit{Gaia DR3} sources to identify any radio emission from them.

To identify the high proper motion radio sources, we have considered a search radius of 11\arcsec{} which accounts for the maximum proper motion (10.39 \arcsec{}/yr for Barnard's star\footnote{Although Barnard's star is not a known radio star, we wanted to account for all possibilities}) and VLASS astrometric offset of 0.5\arcsec{}. Now for all the matches, we obtained the date of observation for each of them from the Subtitle Information Table\footnote{Refer to \href{https://cirada.ca/vlasscatalogueql0}{VLASS Quick Look and Single Epoch Catalogs web page} for details on each VLASS catalog.}. Next, we performed proper motion correction of the individual high proper motion \textit{Gaia DR3} sources to the exact epoch and then re-searched for any VLASS components within $1\arcsec{}$ search radius. This way we recovered 6 high proper motion sources which were not found by general cross-matching.

\begin{figure*}
  \centering
  \includegraphics[width=0.3\textwidth]{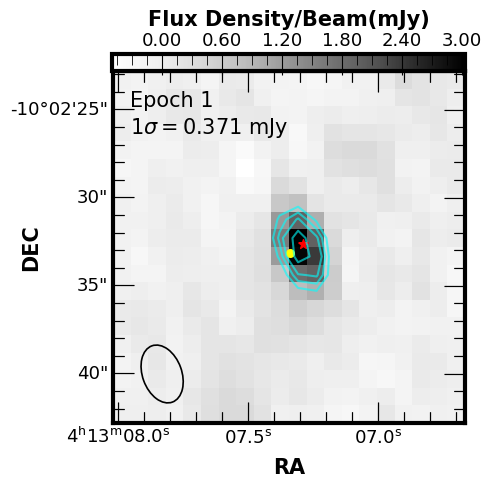}
  \includegraphics[width=0.3\textwidth]{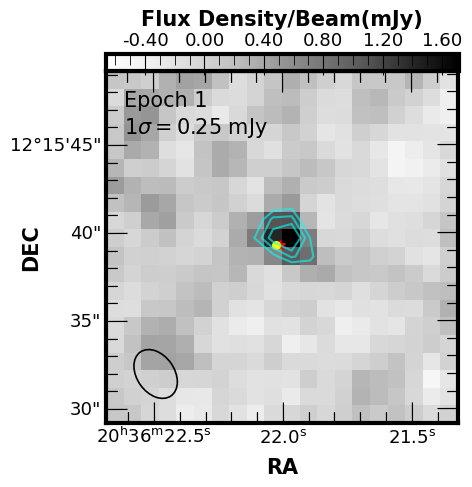}
  \includegraphics[width=0.3\textwidth]{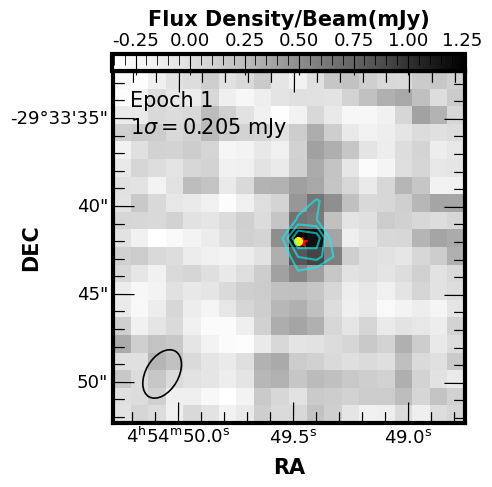}
  \includegraphics[width=0.3\textwidth]{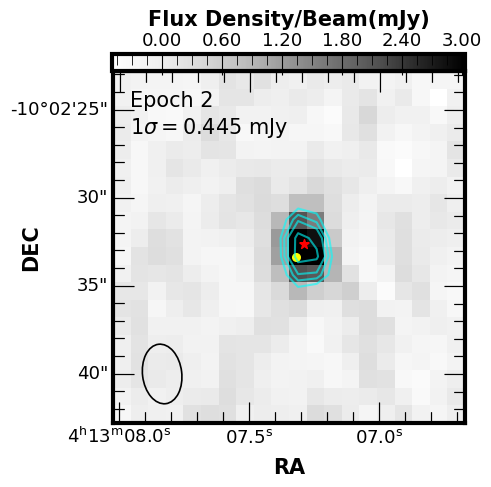}
  \includegraphics[width=0.3\textwidth]{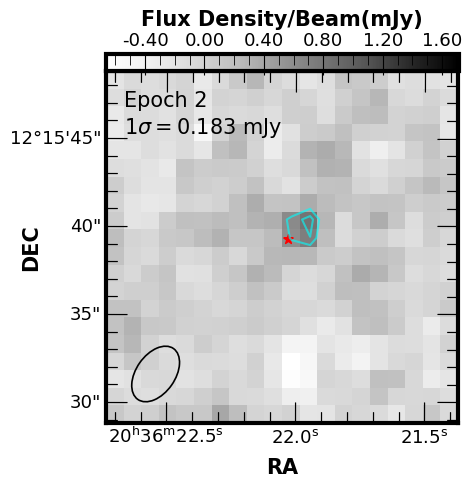}
  \includegraphics[width=0.3\textwidth]{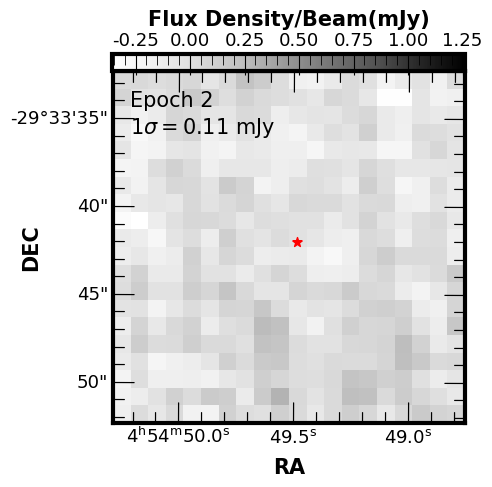}
  \caption{Cutouts\footnote{\href{http://cutouts.cirada.ca/}{CIRADA cutout service}} of Epoch 1 (\textit{top panel}) and 2 (\textit{bottom panel}) observations of 1 source from each of the 3 different categories: Confirmed Radio Sources (Gaia DR3 3192038390381997056) on \textit{left panel}, Candidate Radio Sources (ASAS J203622+1215.3) on \textit{middle panel}, Candidate Transients or highly Variable Sources (ST Cae) on \textit{right panel}. Contours are drawn at $3\sigma$, $4\sigma$, $5\sigma$ and $7\sigma$ levels where $\sigma$ is local rms noise. The synthesized beam at FWHM of the fitted component is shown in the lower-left corner of the images. The same color scale is used for both epochs to highlight the variability in source signal and local noise. The red star marker shows the epoch corrected \textit{Gaia DR3} positions and the yellow marker shows the VLASS coordinate as reported in the QL catalog.}
        \label{category_images}
\end{figure*}

\subsection{Additional filtering of the data}\label{filter}

The Canadian Initiative of Radio Astronomy Data Analysis (CIRADA) has used certain flags on the data in QL1 and QL2 catalogs to denote the quality of the data. We have used the following constraints on our sample:
\begin{itemize}
    \item Images in the quicklook catalog (sub-tiles) have overlapping patches with each other. Components\footnote{Distinct localized regions of emissions in a radio image detected by source detection algorithms (PyBDSF for VLASS) are referred to as components.} detected in these overlapping areas have been cataloged twice. CIRADA has identified these duplicates and flagged them `0' for unique components, `1' and `2' for brighter and fainter duplicate components respectively. We only retain components with duplicate flags `0' or `1'.
    \item CIRADA detects components in the images by detecting blobs (flux islands) and then fitting Gaussian to the flux islands. They flag components where a blob has been detected but no component has been fitted as ``Empty flux islands". These have been denoted in the catalog with \textit{$S\_Code = E$}. We remove such components from our main sample of cross-matches and investigate the images individually.
    \item We only retain components with quality flag, $QualFlag =(0|4)$. This ensures that we do not have detections which have peak flux density lower than 5 times the local rms, or detected components which are side-lobe features (artifacts) of nearby bright sources.
\end{itemize}

\section{Results}\label{results}

\begin{figure*}[t]
    \centering
    \includegraphics[width = 6.3in]{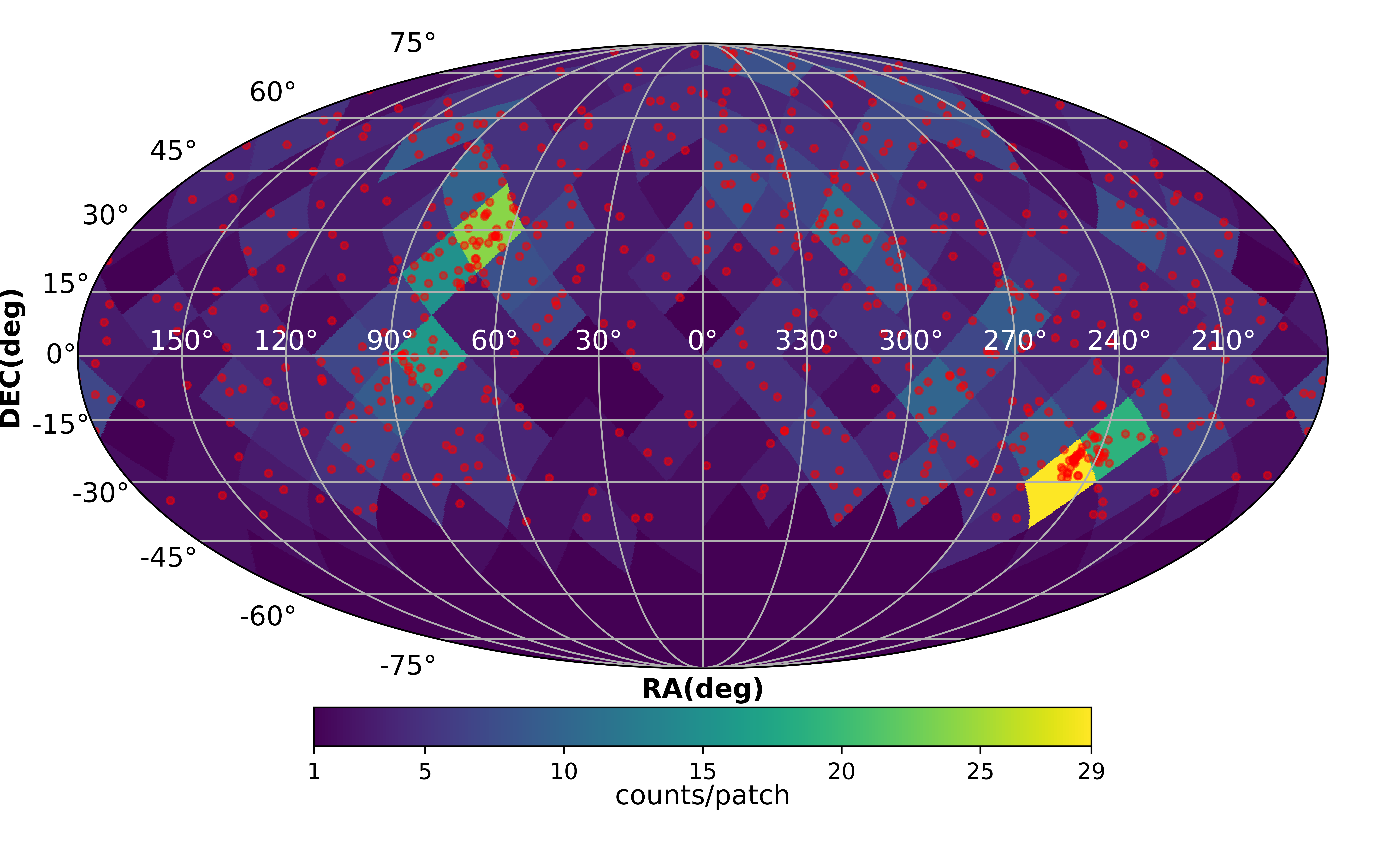}
    \caption{Distribution of all the confirmed and candidate sources on the sky. This color map on a Molleweide projection shows the source density per resolution and red dots are the source coordinates.}
    \label{fig: map}
    \end{figure*}

Putting constraints on the \textit{Gaia DR3} catalog (parallax precision and distance cut-off), we obtained about 16 million sources which were then cross-matched with $\sim$ 3.3 million and $\sim$ 3 million sources each in VLASS QL1 and QL2 catalogs, respectively. The cross-match strategies yield 564 epoch 1 and 563 epoch 2 components. Additional filtering of these samples by applying VLASS flags mentioned in Section \ref{filter} leaves us with 417 epoch 1 and 415 epoch 2 components in Sample A and B respectively. We then checked for common components in Sample A and B for robust source identification. However, upon inspecting the image tiles, we found that some of the detected components were not cataloged. The reasons for this ambiguity, and the methods we implemented to bypass it to ensure that we do not miss any source, has been outlined in the next section. 

\subsection{Source classification based on SNR and multi-epoch detection}

Cross-matching samples A and B using the \textit{Gaia DR3}-ID, yielded only 203 common sources. Inspecting the cutout images and image tiles using CASA \citep{casa}, we found additional 190 sources detected with $\geq 5 \sigma$ radio signal in both epochs, but missing from one of the QL catalogs. These sources were found to be missing because of the following reasons (as also listed in the VLASS catalog user guide).
\begin{itemize}
    \item Different component detection algorithms used for QL 1 and 2.
    \item Excessive noise in one of the epochs resulting in low SNR of the source. Consequently, the source was missed by the detection algorithm or the component was flagged. 
    \item Detection affected by the side-lobe of a bright nearby source, hence getting flagged in one of the catalogs.
\end{itemize}
Due to equal median sensitivity in both epochs, VLASS must have observed any quiescent (steady) radio emission in both epochs. The epochs are well separated by 32 months to also be able to detect variability with large time scales.  Therefore a source detected in one of the catalogs can be missing from the other catalog because of variability or transient emission. Further, there might be some erroneous component fitting to background artifacts. We manually inspected image tile of all the components in Sample A and B using CASA to robustly determine the steady and variable radio objects, candidate transient radio objects, and false detections.

Based on detected signal to noise ratio and epoch-to-epoch variation, we could classify our sample into three types of sources: 
\begin{enumerate}
    \item Sources detected in both epochs (SNR $\geq 5\sigma$)
    \item Sources detected in one epoch and only marginally detected in the other ( $4\sigma <$ SNR $< 5\sigma$)
    \item Sources detected only in one of the epochs, missing from the other (SNR $< 4\sigma$)
\end{enumerate}

The three categories have been discussed in detail in the subsequent sub-sections and illustrated using cutout images of three distinct type of sources as examples in Figure \ref{category_images}.

\subsubsection{Confirmed Radio Sources}

The robust sample selection method outlined in previous sections returned 391 single matches and 2 double matches between VLASS QL1, QL2 (both epoch) radio sources and \textit{Gaia DR3} counterparts. They were all detected in VLASS with SNR $\geq 5\sigma$ in both epochs. We group these \textbf{393} sources into \textit{Category 1} which we here onwards refer to as ``confirmed radio sources". However, due to the reasons outlined above, 203 were cataloged on both samples (thus obtained by directly cross-matching samples A and B) and 190 were identified only after further inspection of the image tiles. These 190 sources were either missing from the quicklook catalogs or flagged. We used CASA software to measure the fluxes for the missing sources and added them to the category of confirmed radio sources. The left panel of Figure \ref{category_images} demonstrates an example of one such source.

\subsubsection{Candidate Variable Sources}
Similar to confirmed sources, after cross-matching and image inspection, we could associate 14 of the radio sources to a Gaia counterpart in both epochs. However, the detections of these sources in one of the epochs are only tentative ($4\sigma <$ SNR $< 5\sigma$). We found 14 such sources and grouped them into \textit{Category 2} and refer to them here onwards as ``candidate variable sources". Follow-up observations or the ongoing (at the time of writing this article) 3rd epoch VLASS observation is necessary to confirm emission from these candidate variable radio sources. The middle panel of Figure \ref{category_images} demonstrates an example of one such candidate variable source.

\subsubsection{Candidate Transient objects}
Significant ($F_{peak}>5\sigma$) emission in one of the epochs but no detectable emission in the other epoch can mean either of the following:
\begin{itemize}
    \item These are highly variable sources.
    \item These are transient sources which were emitting during one of the epochs.
    \item These are artifacts. 
\end{itemize}
We found 191 such sources. Since we cannot confirm the nature of their emission without further observations, we grouped them into \textit{Category 3} and refer to them here onwards as ``candidate transients". The right panel of Figure \ref{category_images} demonstrates an example of one such ``candidate transient source".

Thus to summarise the above results: we cross-matched samples A and B, inspected image tiles of individual components, and obtained a total 603 VLASS sources that have a \textit{Gaia DR3} counterpart. We found 391 confirmed radio sources, 198 candidate transients and 14 candidate variable sources. These include a total of 11 high proper motion sources, 6 of which were not found using general cross-match strategy. Three of them are confirmed radio sources and 3 are candidate transients. Only 2 VLASS components got cross-matched to double \textit{Gaia DR3} sources - one of them is associated with a known spectroscopic binary (SB) system \textbf{HD 239702}  \citep{SB*} and the other component is associated with a close double. Upon NED coordinate query, we found some sources in our sample to be of extra-galactic origin. Further investigating the individual images of the remaining sources, some of them had extended double lobed AGN-like features. We discarded 24 such radio detections which are likely of extra-galactic origin or PyBDSF artifacts.

\subsection{Source distribution in the sky}

\begin{figure}[h]
  \centering
  \includegraphics[width=0.48\textwidth]{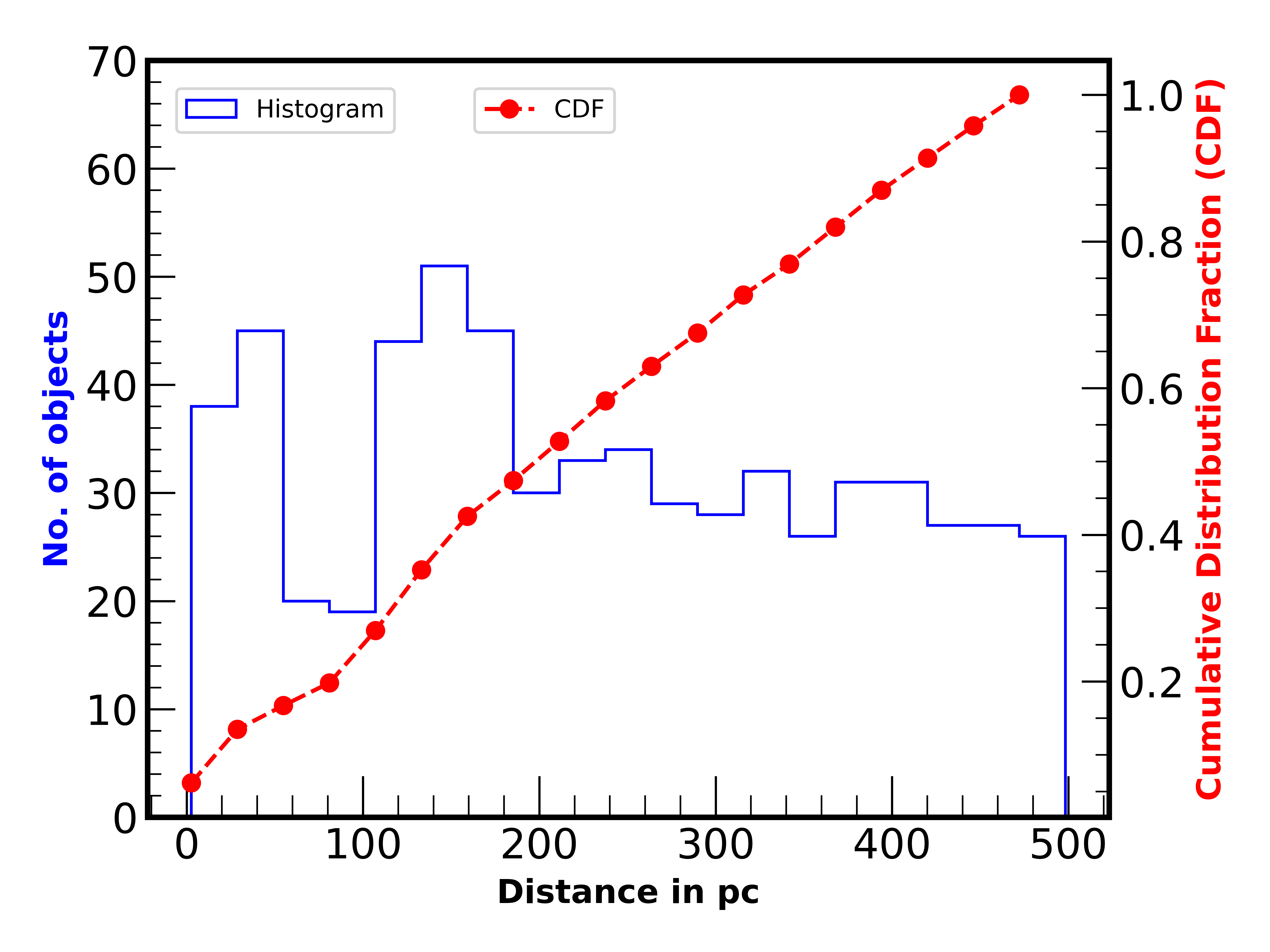}
  \caption{Distance distribution of the entire sample. The solid blue line represents the distribution of distances to all the sources. The cumulative distribution function is shown in red.}
  \label{dist}
\end{figure}

We have shown the sky distribution of all confirmed radio sources, candidate variable sources and candidate transients in Figure \ref{fig: map}. Using Heal\textit{py}, the map was pixelated and the color-map for the source density distribution was created using 2-D histogram. The map shows a distinctively dense patch of the sky of $\approx 10^{\circ} \times 10^{\circ}$ area containing 43 sources in the southern celestial hemisphere near the Ophiuchus molecular cloud. There are two more dense regions, one in the northern sky and the other near the celestial equator containing 21 sources in the Orion Molecular Cloud (OMC). These dense patches contain confirmed and tentative detections from several YSOs, binaries and variable objects. A serendipitous wide-field survey of these patches monitored over a long time can confirm the candidate sources and provide insights about this population of young radio stars. Figure \ref{dist} shows the distribution of distances to all the 603 radio sources with Gaia DR3 counterparts. 

\begin{figure*}[t]
    \centering
    \includegraphics[width = 6in]{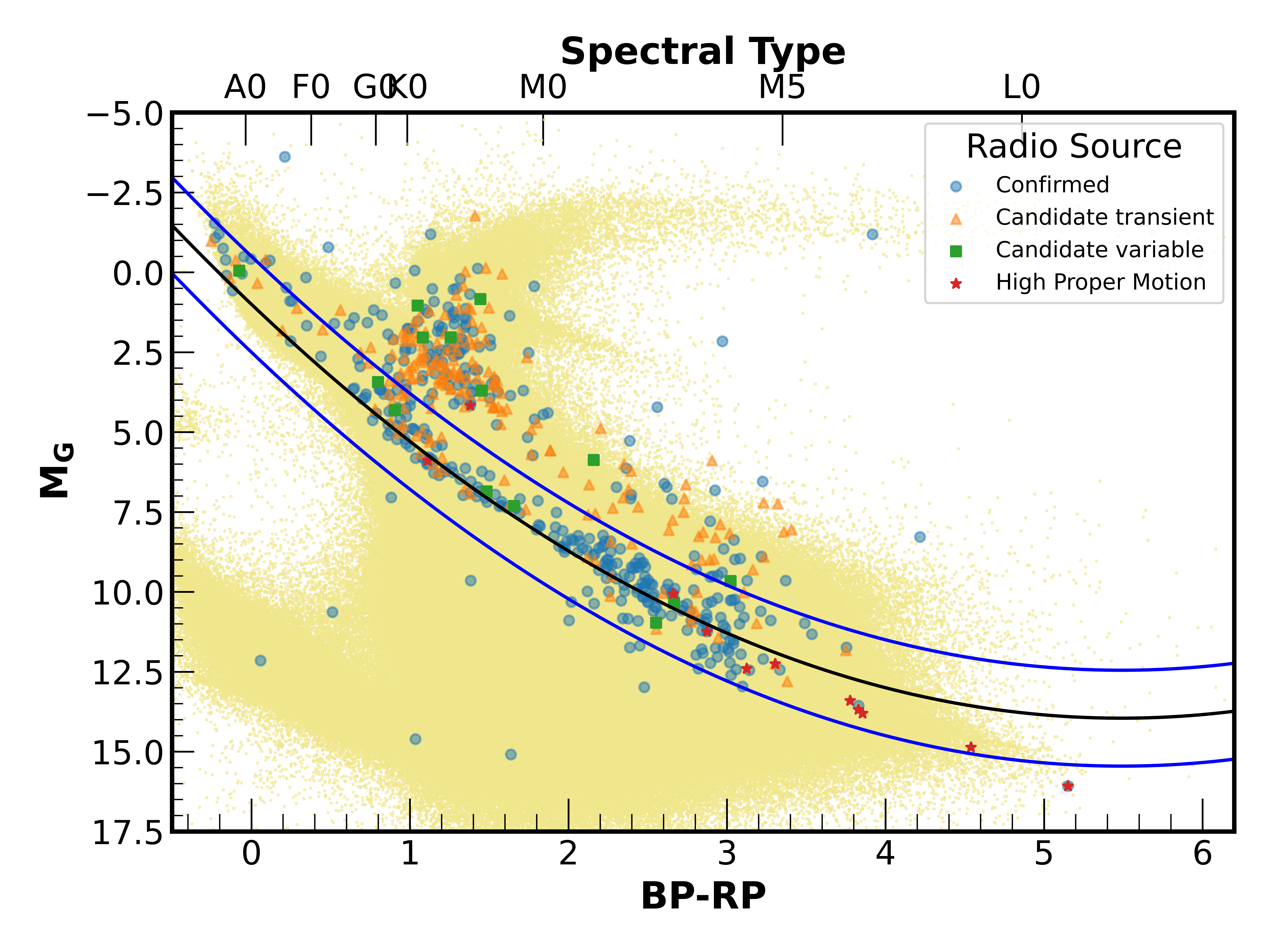}
    \caption{The yellow scatter plot in the background shows all the \textit{Gaia DR3} optical sources within 500~pc. The foreground scatter plot is for 3~GHz sources in our total sample. The 3 categories and stars with high proper motion (PM $> 0.5\arcsec{}/year$) are are shown in different markers. The black line corresponds to a fit for the MS sources in the CMD (Equation \ref{MS}). The blue lines represent the tolerance on $M_G = \pm 1.5$ for the MS sources.}
    \label{fig: HR Diagram}
\end{figure*}

\subsection{Gaia color-magnitude Diagram}\label{HRD}

\textit{Gaia DR3} photometry provides us with apparent magnitudes. Using $BP-RP$ and $Gmag$ from the \textit{Gaia DR3} catalog, and the main sequence cutoff from \cite{bihan2024host}, and Narang et al. (under review) based on the main sequence color-magnitude relation from   \citep{mamajaek}\footnote{\href{https://www.pas.rochester.edu/~emamajek/EEM_dwarf_UBVIJHK_colors_Teff.txt}{Updated Table based on Pecaut and Mamajaek 2013}}, we categorized the 3~GHz radio sources that we obtained from the VLASS-\textit{Gaia DR3} cross-match into different spectral types. To separate the main-sequence objects in the color-magnitude diagram, we implemented the following equation,
\begin{equation}\label{MS}
    M_{G} = -0.43(BP-RP)^{2} + 4.72(BP-RP) + 1
\end{equation}
Figure \ref{fig: HR Diagram} shows the color-absolute magnitude plot (CMD) for our entire sample. To highlight the contrast in the population of radio sources against the optical sources, we have plotted the VLASS sources in the foreground of the \textit{Gaia DR3} sources (yellow).


\subsection{Distinct populations of radio emitters}\label{obtypes}

\begin{figure*}[t]
    \centering
    \includegraphics[width = 6in]{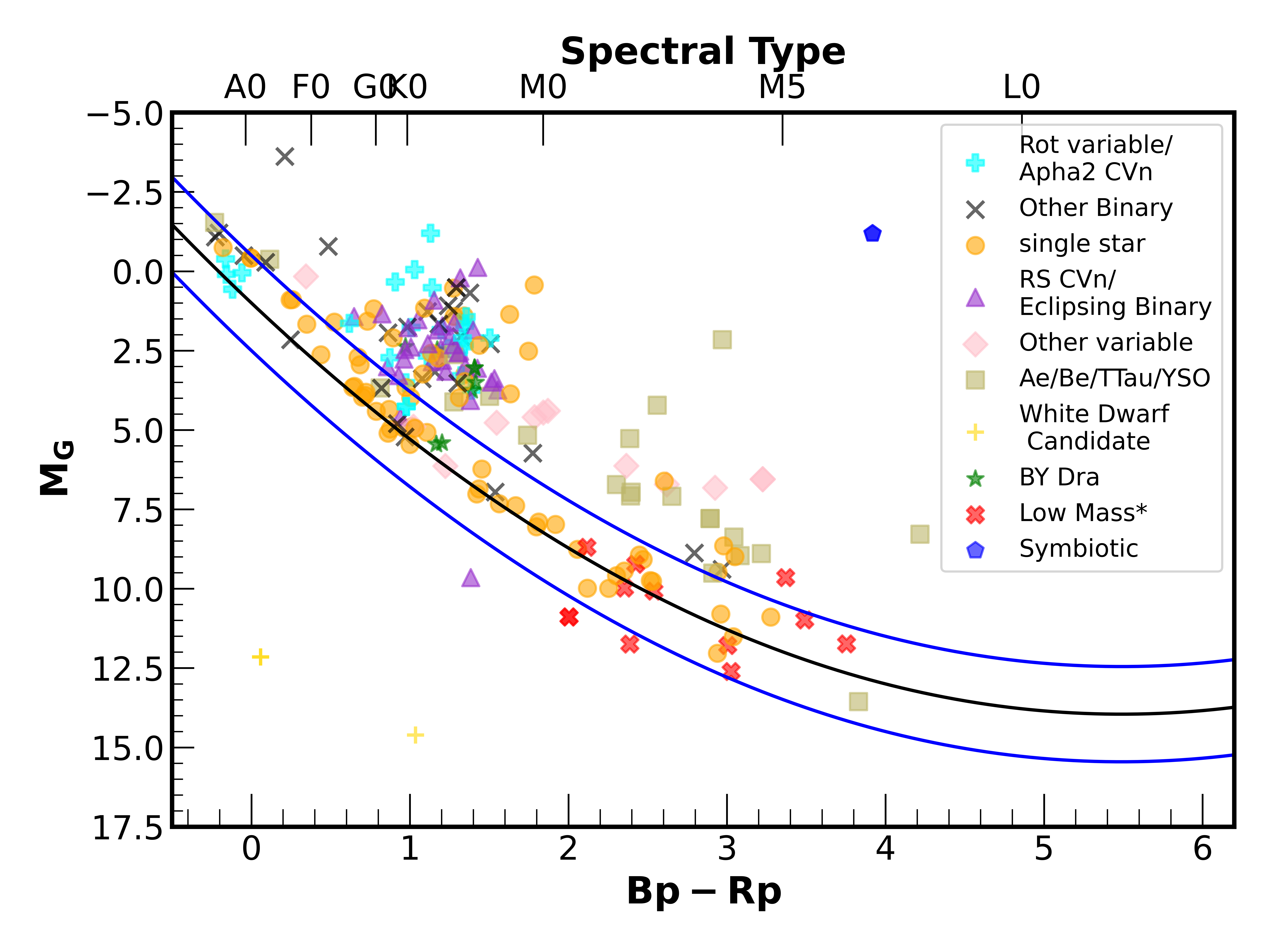}
    \caption{The \textit{Gaia} CMD for 207 of the 391 confirmed radio sources that were also found on SIMBAD by a $2.5\arcsec{}$ radius query. Different object types have been represented using different shapes.}
    \label{fig: Radio HR Diagram}
\end{figure*}

Prior knowledge of the object type combined can provide us insights into the physical processes driving the radio emission in these sources. Following \cite{vlass_new_hv, pecRX_hv} we used data from SIMBAD  \citep{simbad} to classify the object type for the detected sources. Among the 391 confirmed radio sources, we could only find 207 sources on SIMBAD. Figure \ref{fig: Radio HR Diagram} shows the Gaia CMD for the SIMBAD identified confirmed radio sources.

From Figure \ref{fig: HR Diagram} and Figure \ref{fig: Radio HR Diagram}, we can already infer some of the demographic features.

\begin{itemize} 
    \item About $4\%$ of the 603 sources (i.e., 23) in our entire sample are B, A and early F-type radio stars. 
    \item Roughly $48\%$ of the radio sources lie above the main sequence. These sources are mostly binaries and YSOs as seen in Figure \ref{fig: Radio HR Diagram}. These sources are highly variable radio sources. A lot many of them are candidate variables and candidate transients.
    \item late F to early M type typical radio stars constitute $\approx 27\%$ of our sample. They are mostly single stars and are expected to emit in radio because of chromospheric activity.
    \item Late M dwarf types constitute $\approx 19\%$ of our sample despite the fact that they do not possess magnetic dynamos. Studying radio emission from ultra-cool dwarfs will help us bridge the gap in our understanding of origin of stellar and planetary magnetic fields.
    \item 2 out of the 203 SIMBAD-identified sources in our sample are white dwarf candidates. 
\end{itemize}

In section \ref{discussion}, we discuss how various properties of radio emission vary across spectral types and object types. Some of the known radio emitters from these SIMBAD-identified object types have been discussed in the Appendix \ref{sources}. 

\section{Discussion}\label{discussion}

By studying the nature of radio emission, we can estimate the region of origin (photosphere, chromosphere, corona, wind, etc) and the physical processes driving the emission (photospheric magnetic fields, chromospheric activity, coronal ejections, magnetically trapped winds, etc), which provides insights into the stellar structure and ambient conditions. We discuss the radio brightness temperature, radio variability and correlation between radio and X-ray flux for our sample and how do they vary across different spectral and object types. In-band spectral index calculations are done using single epoch images which can be found in the SE catalogs. However, they are partially incomplete, and therefore, we do not analyse spectral index variation in our study.

\subsection{Brightness Temperatures}\label{Tbright}
The temperature of a blackbody having the same observed radio brightness (specific intensity, $I_{\nu}$) at a frequency $\nu$ as observed in a source is referred to as the brightness temperature ($T_b$) for that source at $\nu$.  Brightness temperature is a proxy for the the emission mechanism (coherent or incoherent; bremsstrahlung, gyrosynchrotron, ECMI, etc) \citep{gudel_review}. Since the continuum at radio frequencies can be described by Rayleigh-Jeans Law, specific intensity $I_\nu$ is given as,
$$
    I_\nu = \frac{2kT_b\nu^2}{c^2}
$$
where $k$ is the Boltzmann's constant and $c$ is the speed of light. The observed flux density, $S_\nu = I_\nu A/d^2$ where A is the cross-sectional area of the source perpendicular to the line of sight and $d$ is the distance to the source. Now we can write the brightness temperature as \citep{gudel_review},

$$
T_b = \frac{S_\nu d^2 c^2}{2Ak\nu^2} \label{eqtemp_a} \\
$$
\begin{equation}\label{temp_b}
T_b \sim \left (\frac{S_\nu}{1~mJy}\right)\left (\frac{d}{1~pc}\right)^2\left (\frac{10^{11}cm^2}{r^2}\right)\left (\frac{1GHz}{\nu}\right)^2 10^7K 
\end{equation}

where we assume a perfectly spherical source region of radius $r$. For our sample, $\nu = 3~GHz$ and $S_{3GHz}$ is the peak flux density ($Fpeak$) as provided by the quicklook catalog or inferred using CASA (for components that were not cataloged). The QL catalogs provides the deconvolved beam size and peak and integrated flux densities. Since we are dealing with point sources, we consider the listed peak flux density per beam for the analysis. We calculated brightness temperatures for sources whose stellar radius $r$ is obtained from \textit{Gaia DR3} catalog for Astrophysical Parameters \citep{gaia_dr3}. These radii estimates are based on General Stellar Parameterizer using photometry (GSP-Phot) that uses certain forward modelling approaches as explained in \cite{GSP_Phot}.

\subsection{Emission mechanism}
VLASS is not sensitive enough to detect photospheric continuum radio emission. The hottest star in our sample is a B8V type. It has to be within 6~pc, for its continuum radio emission to be detected by VLASS, whose RMS sensitivity is $\approx 120~\mu Jy$. The only radio source in our sample within $6~pc$ is an M6V type flaring dwarf \textbf{WX UMa}, which has too cool a photosphere to be detected by VLASS. This ensures that emission from all objects in our sample is of non-photospheric origin.

Other types of emission mechanisms, both thermal and non-thermal, are produced by highly energetic charged particles. Emission mechanisms can be characterized roughly based on brightness temperatures (as described in the review by \cite{gudel_review}):
\begin{enumerate}
    \item \textit{\textbf{Bremsstrahlung radiation}} originates from thermal or relativistic plasma. They are limited up to $T_b \approx 10^6K$. 
    \item \textbf{\textit{Cyclotron radiation}} can result from magnetized thermal plasma. They are limited up to $T_b \approx 10^8K$.
    \item \textbf{\textit{Gyrosynchrotron emission}} can occur due to mildly relativistic thermal as well as power law electron distributions. They can explain $T_b$ up to $\approx 10^9K$
    \item \textbf{\textit{Synchrotron radiation}} can be caused only by relativistic power law electron distribution. They are highly polarized and can go as high as  $T_b \approx 10^{12}K$.
    \item \textbf{\textit{Coherent plasma radiation}} are also highly polarized and can explain emissions with $10^{12}K < T_b < 10^{16}K$.
    \item \textbf{\textit{Electron cyclotron maser}} accounts for any emission with $T_b > 10^{16}K$. They are almost completely polarized radiation. 

\end{enumerate}

\begin{figure}[h]
    \centering
    \includegraphics[width = 0.48\textwidth]{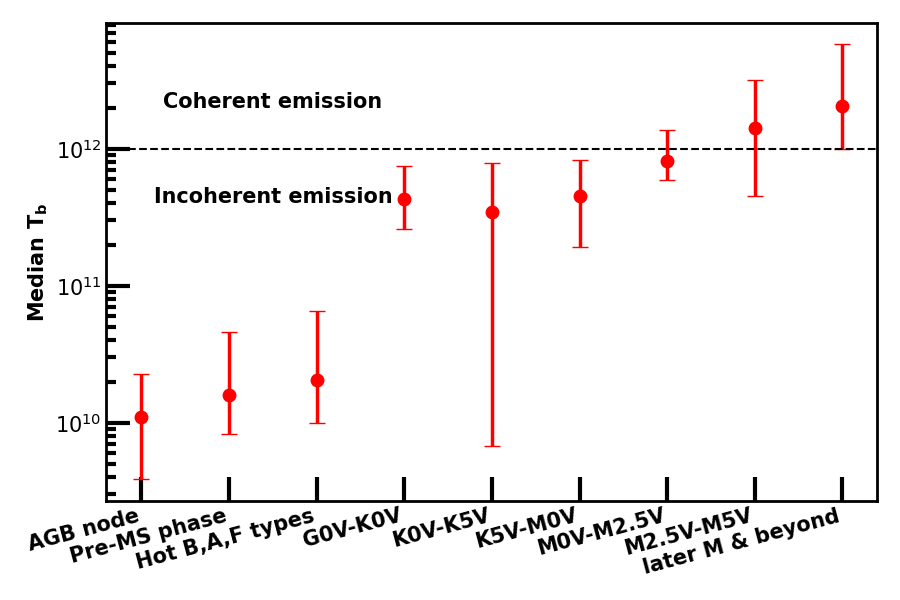}
    \caption{Median brightness temperatures at 3~GHz of giant, pre-main sequence and different main sequence spectral types. Quartile 1 and 3 have been shown using error-bars which represent the central scatter around the median. Coherent and Incoherent emissions can be distinguished by the $T_b = 10^{12}K$  \citep{gudel_review} line.}
    \label{fig: Brightness Temprature}
\end{figure}

Figure \ref{fig: Brightness Temprature} shows the median of brightness temperatures calculated using Equation \ref{temp_b} of different populations of radio sources in our sample as identified from the Gaia CMD.

There is a known underestimation of $8-15 \%$ in the peak flux measurements as reported by CIRADA in the QL1 catalog. Further, the radius of the stars provided in \textit{Gaia DR3} can give the best estimate of $T_b$ only if the emission is of chromospheric origin (assuming the radius of a chromosphere is similar to the  photosphere). It would, however, underestimate $T_b$ if the emission is a collimated beam from a smaller patch of the star ($r_{source}<r$). It would overestimate $T_b$ if the emission is from the entire corona, disc or wind which has $r_{source}>>r$. Therefore, our calculated $T_b$ of individual objects alone cannot provide us with the most precise picture of the actual origin of their radio emissions. Nevertheless, from \autoref{fig: Brightness Temprature}, we can draw a general trend in the brightness temperatures of different object types and make general inferences, which can be tested with follow-up observations.  \\

From Figure \ref{fig: Brightness Temprature}, we observe that:
\begin{itemize}
    \item Incoherent gyrosynchrotron emission ($T_b<10^{12}K$) mechanisms dominate in the objects above main sequence (MS) and hot B,A and F type stars.
    \item G and K type stars are mostly incoherent gyrosynchrotron emitters with relativistic power law electron distributions \citep{gudel_review}.
    \item In later K-type and early M-type stars, relativistic synchrotron emission from plasma bursts and flares dominate the radio emission \citep{calingham,vlass_new_hv}.
    \item Main sequence stars show a general trend of increasing brightness temperatures for cooler stars, along with a transition in the emission mechanism from incoherent to coherent around mid-M type (see Figure \ref{fig: Brightness Temprature}).
\end{itemize}

Although our analysis indicates a trend of incoherent emissions from B and A type stars because of larger photospheric radius, multi- frequency radio observations of hot stars have revealed highly polarized ECM emission beamed from smaller source regions \citep[e.g.,][]{Das_2020, Das_2022}. Therefore, observed B and A type stars need to be followed up at other frequencies to unambiguously interpret their radio emission. We also have 4 main-sequence F0-F6 type stars in our sample. Since early F-type stars have shallow convective zones and weak stellar wind, we do not expect them to be able to power detectable radio emissions. Therefore further follow up on these F-type stars can reveal new insights into stellar structure (Ayanabha et al. in preparation).

Cooler dwarfs (later than M2.5V types) are found to exhibit high brightness temperatures ($T_b > 10^{12}K$) indicating that coherent emissions are ubiquitous in ultra-cool and brown dwarfs. This spectral range corresponds to the transition from partial to fully convective interiors \citep{reiners, Baraffe_2018,Jao_2018}. With the lack of tachocline, magnetic fields in these type of objects is thought to originate from mechanisms different from solar dynamos \citep{Kao17, vlass_new_hv} which possibly powers the observed radio emission. If these dwarfs possess weaker magnetic fields, they might emit mostly at much lower frequencies. We should be able to observe faint decimetric emission and bright emission at low frequencies  \citep{calingham,vlass_new_hv} from these dwarfs. Low frequency observations are crucial to completely understand the nature of these radio emitting dwarfs (\cite{Burningham}, Narang et al. in preparation). The dearth of detection of late M-types and cooler dwarfs in our sample is because they are faint emitters at 3~GHz and VLASS single epoch quicklook images do not have the required sensitivity to detect them. 

\begin{figure}[h]
    \centering
    \includegraphics[width = 3.3in]{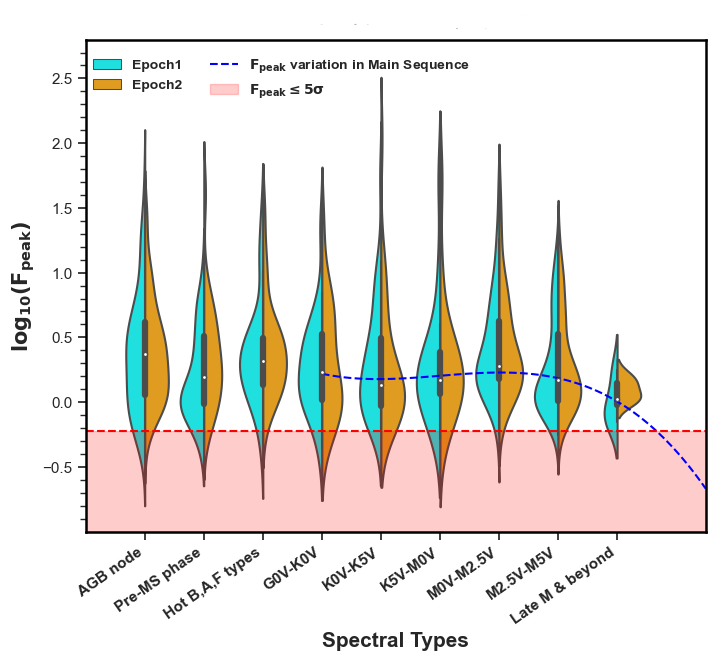}
    \caption{Variation of peak flux density with spectral types: White points and bold boxes inside the violins show median and inter-quartile range while the violin edges represent the distribution. Single epoch VLASS images are mostly sensitive to the area above the shaded region ($F_{peak} > 5\sigma$). The blue dashed line shows extrapolated median flux variation for late M types.}
    \label{violin}
\end{figure}

\autoref{violin} shows that observed median flux densities drop off for cooler dwarfs. Emission from these dwarfs are possibly due to ECMI which is beamed from small patches \citep{Toet_2021, Vedantham_emission}, resulting in the net radio output being relatively low. Deeper observations with higher signal-to-noise ratio or even the 3 epochs stacked deep cleaned VLASS images are likely to reveal more cool dwarfs emitting in 3~GHz.

\subsection{Radio Variability of 3~GHz emitters}

\begin{figure}[h]
    \centering
    \includegraphics[width = 0.48\textwidth]{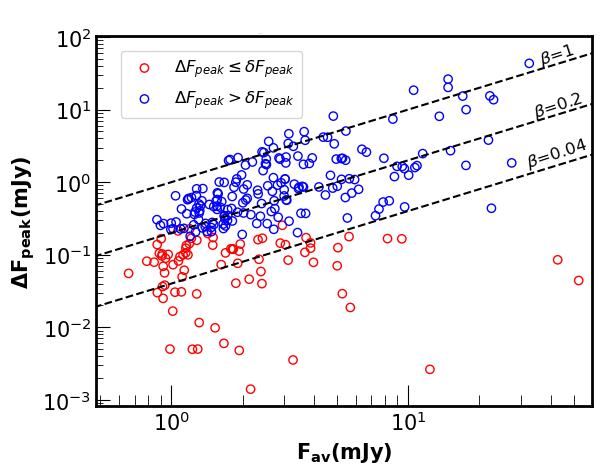}
    \caption{Source varying by more than the uncertainty in flux measurement ($\delta F_{peak} = \sqrt{\delta F_1^2 + \delta F_2^2}$) is denoted by blue and otherwise by red. The dashed lines indicate different values of $\beta$}
    \label{fig: variability}
    \end{figure}

Non-thermal radio emissions are generally variable whereas thermal emissions are quiescent \citep[e.g.,][]{gudel_review}. We have used the two epoch flux measurements for confirmed sources to report a general trend of variability for our sample. We have used the quantity $\beta = \Delta F_{peak}/F_{av}$ to quantify the extent of variability as visualised in Figure \ref{fig: variability}. $\Delta F_{peak} = |F_1 - F_2|$ is the change in peak flux measured between two epochs and $F_{av} = (F_1 + F_2)/2$ is the mean peak flux. Any source with $\beta > 0.2$ should be considered to be significantly variable (between the two epochs). Otherwise, the radio emission can be\footnote{Short-term variability caused by flares or periodic variation triggered by SPIs can be missed by on the fly short integration VLASS observations.} considered to be quiescent.

 It is to be noted, however, that two epochs of observations are not sufficient to provide a true measure of the amplitude and timescale of the variability. VLASS Epoch 3 data release and individual follow-ups are necessary for properly characterizing  short-term and even long-term variability.
    
\subsection{Correlation between Radio \& X-ray output}
Flares in stars are generated when energetic plasma trapped in magnetic loops is released. Most of the energy in the magnetic fields is emitted in low-frequency X-rays through thermal Bremsstrahlung radiation and is thought to be responsible for coronal heating while a fraction of the energy accelerates the trapped plasma \citep{gudel&benz}. This non-thermal power law plasma emits gyro-synchrotron radio emission. Therefore, we expect a correlation between observed radio and soft X-ray emissions from magnetic plasma environments present in stellar atmospheres \citep{gudel93, benz_gudel}. Such correlation is indicative of the contribution of flares in heating of stellar corona and efficiency with which flares accelerate the plasma.
On the other hand, neutral atmospheres in cooler dwarfs are quieter in X-ray but can still be radio bright. They are expected to deviate from such a Radio-X-ray correlation \citep{Burningham, pecRX_hv}.

\begin{figure}[h]
    \centering
    \includegraphics[width = 0.48\textwidth]{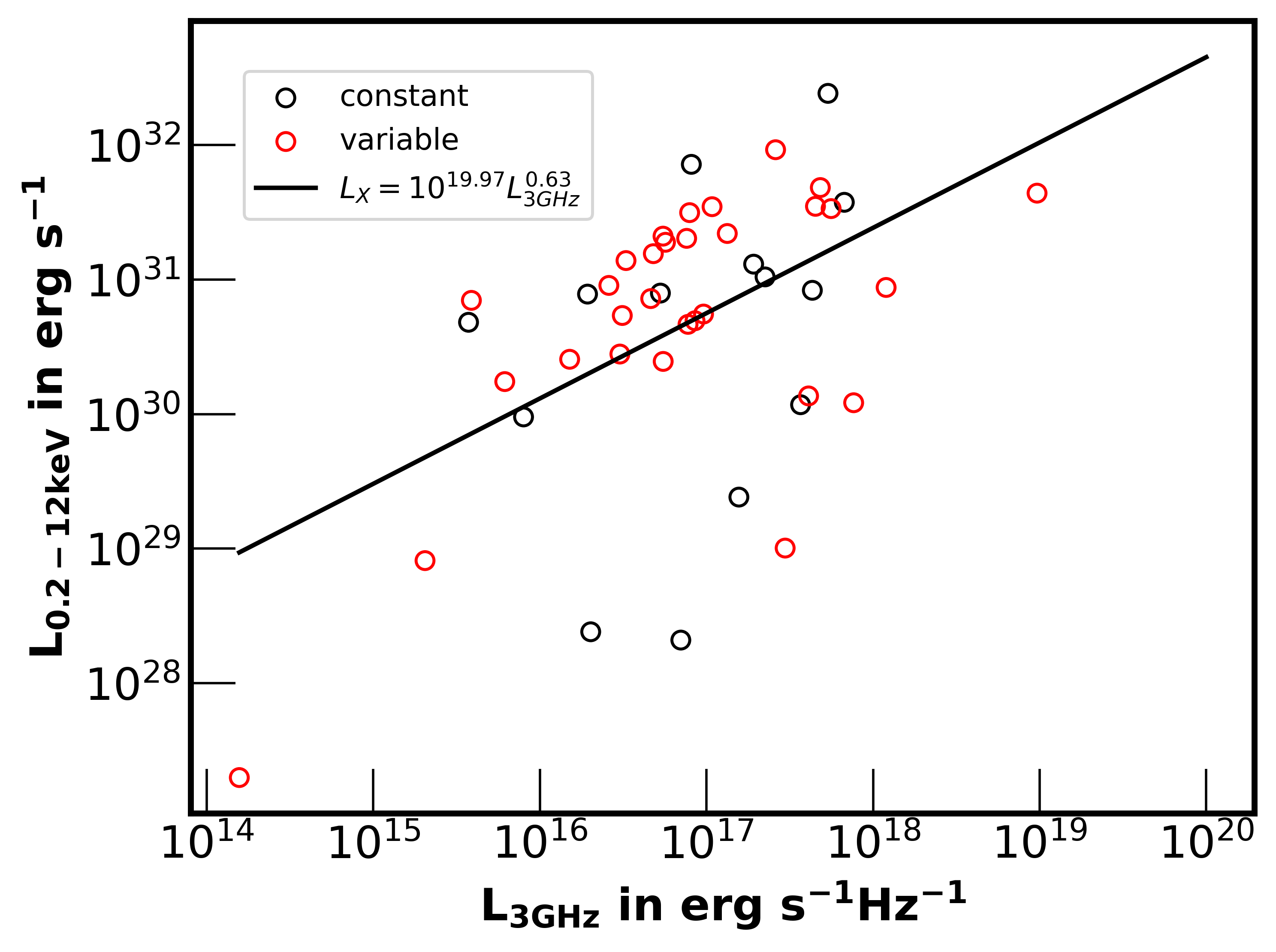}
    \caption{Variable ($\beta > 0.2$) and quiescent ($\beta < 0.2$) sources are denoted by red and black points respectively. The black line is a linear fit.}
    \label{fig: GB}
\end{figure}

We cross-matched our sample of confirmed radio sources with the XMM Newton Serendipitous Survey Catalog \citep{webbxmm, xmmdata} to obtain 44 matches. Their X-ray (0.2-12~keV) and Radio (2-4~GHz) luminosities (derived from average peak flux density measurement), have been plotted in Figure \ref{fig: GB}.

To test the correlation between radio and X-ray log luminosity, we opted for the Pearson-$\rho$ test rather than a rank correlation test since we do not have a very large number of rank ties. The Pearson test gives $\rho$ = 0.5 with a p-value of $5.3\times 10^{-4}$ indicating a positive correlation with moderately significant confidence. For the entire matched sample, we find a best fit correlation as, 
\begin{equation}
    L_X = 10^{19.97\pm2.75}L_{3GHz}^{0.63\pm0.16}
\end{equation}

Follow-up near-simultaneous observations at X-ray and radio wavelengths of these types of sources is crucial to precisely study this correlation.

This correlation relates stellar chromospheric activity to the heating of stellar coronae, which is not very well understood. Also, this data corroborates the mounting evidence of the fact that flares are not uncommon in young stellar objects. Understanding their origin is key to understanding the origin of stellar magnetic fields and the loss of fossil fields.

\section{Summary}\label{summary}

In this work we carried out a systemic analysis of the VLASS epoch 1 and 2 data, in combination with \textit{Gaia DR3} to investigate the population of radio stars within 500~pc. \textit{VLASS Quick Look (QL1 and QL2)} catalogs were crossmatched with \textit{Gaia DR3} to identify 3GHz radio stellar sources within 500~pc. To reduce chance-coincidence with background artifacts and galaxies and to ensure that we do not miss any radio source due to inaccuracy in source detection algorithm, we implemented robust cross-matching techniques and inspected individual image tiles. Below, we summarize the major results from our demographic analysis of a homogeneous radio stellar population that we produced.

\begin{itemize}
    \item Epoch 1 detections were compared with Epoch 2 detections to classify the radio sources into three categories:
    \begin{itemize}
        \item \textit{Confirmed radio sources}: They were detected in both epochs with $\geq 5 \sigma$ signal.
        \item \textit{Candidate variable sources}: They were detected in one epoch with $\geq 5 \sigma$ signal but only tentatively ($4\sigma <$ SNR $< 5\sigma$) in the other epoch. 
        \item \textit{Candidate transient sources}: They were detected with $\geq 5 \sigma$ signal in one epoch but no detection in the other. 
    \end{itemize}

    \item Based on the Gaia color-magnitude diagram we find that most of our sample consists of single as well as binary F, G, K, and M type main-sequence stars. We also report few atypical B, A and early F type radio stars and white dwarf candidates. 
    \item Apart from MS stars we also find a large population of YSOs and moving group members.
    \item We analyse the radio brightness temperature of the sources to infer qualitatively the type of mechanism that drive radio emission in a star of a particular spectral and object type. We find that M dwarfs mostly exhibit coherent radio emission whereas most other stars are typically incoherent radio emitters.
    \item We cross-matched our sample with XMM Newton Serendipitous Survey catalog to test for any relation between radio and X-ray flux. We find a moderately significant positive correlation, $\displaystyle L_X \propto L_{3GHz}^{0.63\pm0.16}$.
    
\end{itemize}

The results from this study and our sample of radio bright stars within 500 pc can serve as a valuable resource for not only understanding the radio emission from stars but also as a robust sample for further follow-up studies. 

\section{Acknowledgement}
We thank the reviewer for their constructive comments, which helped to improve the clarity and sharpness of the presentation. This research has made use of the SIMBAD database, operated at CDS, Strasbourg, France. The National Radio Astronomy Observatory is a facility of the National Science Foundation operated under a cooperative agreement by Associated Universities, Inc. CIRADA is funded by a grant from the Canada Foundation for Innovation 2017 Innovation Fund (Project 35999), as well as by the Provinces of Ontario, British Columbia, Alberta, Manitoba and Quebec. This work presents results from the European Space Agency (ESA) space mission Gaia. Gaia data are being processed by the Gaia Data Processing and Analysis Consortium (DPAC). Funding for the DPAC is provided by national institutions, in particular, the institutions participating in the Gaia MultiLateral Agreement (MLA). The Gaia mission website is https://www.cosmos.esa.int/gaia. The Gaia archive website is https://archives.esac.esa.int/gaia. This study has made use of data obtained from the 4XMM XMM-Newton serendipitous source catalog compiled by the XMM-Newton Survey Science Centre Consortium.
AD would like to thank the Department of Astronomy and Astrophysics (DAA) at TIFR for hosting them during the coarse of this research. The support and resources provided by TIFR were invaluable to the completion of this work. PKN acknowledges support from the Centro de Astrofisica y Tecnologias Afines (CATA) fellowship via grant Agencia Nacional de Investigacion y Desarrollo (ANID), BASAL FB210003.

\appendix
\section{Source description}\label{sources}

Below we briefly discuss some of the sources whose spectral type and evolutionary stages are robustly determined. The SIMBAD identifier of these sources has been highlighted by bold text.

\subsection{Hot B, A, and early F-type stars}
OB-type stars are rare in the solar neighbourhood as evident in the background \textit{Gaia DR3} source plot in Figure \ref{fig: HR Diagram}. Despite the lack of convective layers, due to their radiative outer layers, they have ionised winds and some retain strong fossil fields  \citep{star_mag}. Gyrosynchrotron and synchrotron emission have been observed from optical thick coronae or winds of these stars or colliding wind of contact binaries  \citep{gudel_review}. Since we have a large volume-limited sample, we do find some late B-type stars in addition to a few A and early F-type stars. We confirm steady radio emission from chemically peculiar Be star \textbf{HD 23478} and single MCP star \textbf{HD 182180}. One of the candidate variable sources \textbf{HD34736} has a strong magnetic field and is suspected to be an interacting close binary system  \citep{HD34736}. Young B type star \textbf{$\mathbf{\sigma}$ Ori E} (part of open star cluster $\sigma$ Orionis) shows variable peak radio flux. It is spinning down due to magnetic braking  \citep{orionis}. This star can provide a peek into the early evolution of stellar structure and magnetic fields. Further, we confirm steady emission from 3 $\alpha^2$ Canum Venaticorum ($\alpha^2$ CVn) variable and 2 rotational variable stars and a double star system.

Bright X-ray source  \citep{tau}, \textbf{HD 28867} exhibits variable radio emission. It is a B9V YSO in the Taurus Auriga star forming region. Known Algol type eclipsing binary \textbf{RZ Cas} shows high variability. We further report confirmed and potential transient detections from 3 more eclipsing binaries, 1 double star, 2 pulsating variables and 6 single stars. Likely because of companions, the absolute magnitude and color indices change shifting them above the main sequence. The confirmed variable source, \textbf{TYC 5366-707-1} might be a magnetic chemically peculiar main sequence (MS) F-type star.

\subsection{RS CVn, Spectroscopic and Eclipsing Binaries }
Binaries and visual doubles with main sequence components deviate from the MS (beyond blue lines in Figure \ref{fig: HR Diagram}) due to extra measured luminosity added by their companions. We find 30 confirmed, 18 candidate transient and 1 candidate variable emission from RS CVn binaries. They mostly populate the region above MS ranging from F to M types. The late-type stars in these systems possess strong magnetic fields resulting in gyrosynchrotron emission  \citep{rscvn1} and heightened chromospheric activity due to interacting magnetic fields causing synchrotron or coherent ECMI polarized emission. Though the former is detectable in decimetric bands, the latter is detectable at much lower frequencies  \citep{Toet_2021} not covered by VLASS.  These systems show very high variability. \textbf{DM UMa} dims by over an order of magnitude\footnote{by the extent of variability, we mean the range of measured 3~GHz flux. The variability time scales can be much shorter than 32 months.} from Epoch 1 to 2. They are bright X-ray sources and show strong CaII triplet and H$\alpha$ absorption features which are other chromospheric activity indicators  \citep{chrom_act}.

8 detections were made from close doubles (or visual binaries). High proper motion source, $\alpha$ For A and B are visually separated by $5\arcsec{}$. Using proper motion search and $1\arcsec{}$ search radius, we can confidently associate the radio emission to the X-ray source $\mathbf{\alpha}$ \textbf{For B}. 15 confirmed, 1 candidate variable and 10 candidate transient spectroscopic binaries were found. 6 confirmed, 1 candidate variable and 11 candidate transient sources were associated with eclipsing binaries. Colliding winds and mass transfer can accelerate charges and drive radio emissions in these binaries.

\subsection{BY Dra, Eruptive and $\alpha^2$ CVn Variables}

BY Draconis type variables (BY Dra variables) are chromospheric variables with in-homogeneous photospheric features caused by strong persistent magnetic fields. They can be fast-rotating young stars like \textbf{BO Mic} or can be late-type dwarfs like \textbf{V402 Hya}. 7 confirmed, 8 candidate transients and 1 candidate variable BY Dra sources were found. They populate a large fraction of the region above MS. Rest is composed of $\alpha^2$ CVn, pulsating and eruptive variables and YSOs.  We find 5 Pulsating variable stars in our total sample, out of which only \textbf{HD 218779} is associated to confirmed radio detection and the other 4 were only detected in one of the epochs. The eruptive variable \textbf{2MASS J21103096-2710513} is an optical flaring M-dwarf  \citep{Erup*} which shows potential transient 3~GHz emission. 

\subsection{Young Stellar Objects}

We also find a large number of pre-MS objects in our radio sample. Due to strong magnetic fields, magnetic braking, accretion, and persistent flaring activity are ubiquitous in protostars. \textbf{HD 200391}, and \textbf{EM* MWC 297} are 2 Herbig Ae/Be stars that exhibit steady and variable emission respectively. There are 10 YSO candidates, 6 confirmed detections and 4 potential transient emissions from known YSOs. T-Tauris are highly common in our sample. 8 confirmed detections, 1 candidate variable and 15 candidate transient emissions from  T-Tauris have been found. The \textbf{T Tau} is found in our sample with variable brightness. Variable 5 GHz emission from candidate transient source, \textbf{V1201 Tau} had been observed using VLBI by  \cite{vlbi_tau}.


\subsection{Partially convective stars ($F3V-M2.5V$ MS stars)}

Stars in the mass range $0.4M_{\odot} \lesssim M_* \lesssim 1.45M_{\odot}$ have radiative cores and outer convective layers. Differential rotation in the convective zone beginning at the tachocline (the boundary between the radiative and convective zone) generates the magnetic field in these stars. Shear and turbulence in the field lines caused by differential rotation drive persistent magnetic activities. \textbf{V815 Her} is a steady emitter in our sample that demonstrates sun-like magnetic fields  \citep{her}. Steady gyrosynchrotron emission from solar-type micro-flares or variable emission of short time scales from solar-type bursts, optically thick bremsstrahlung radiation from solar like coronae is common. Coherent emission due to magneto-ionic plasma oscillations is ubiquitous in these magnetic stars.

The 12 candidate transient and 2 candidate variable sources in this range are binaries, variable stars and T-Tauris stars. The 47 confirmed detections are mostly from isolated single stars and few variables and binaries. Expectantly these confirmed sources mostly show quiescent emission. Only 3 are significantly variable while few are only mildly variable.

\subsection{Fully convective stars and Ultra-cool dwarfs}
M2.5V and later type dwarfs ($M_* \gtrsim 0.4M_{\odot}$) have completely convective interiors with no tachocline. We do not have any clue if such stars can generate magnetic fields of their own. Even if they retain fossil fields, they are slow rotors and wouldn't be able to generate as much activity. However, we find roughly 19\% of our entire sample to be late M dwarfs, four of which are ultra-cool dwarfs (later than M6V type) exhibiting emission with high brightness temperatures. Coherent plasma emission or ECMI can be a result of Jupiter-like large-scale magnetospheric dynamics  \citep{vlass_new_hv}.

Eruptive high proper motion M6 dwarfs \textbf{WX UMa} and \textbf{G 272-61B} were seen to significantly vary over two epochs, meaning they might have been caught flaring in one of the epochs. We performed a cross-match of the total sample with the latest ultra-cool dwarf (UCD) catalog  \citep{ucd_sheet} to find 2 cross-matches. We associate the M8.5V type dwarf \textbf{LSR J1835+3259} to confirmed radio source and M7V type dwarf \textbf{LSR J0510+2713} to candidate variable source.

\subsection{Below Main Sequence}
2 confirmed emissions are from candidate white dwarfs. They show highly variable emissions. Terrestrial planets orbiting close to a white dwarf can induce ECMI emissions  \citep{white_dwarfs}. Eclipsing binary system \textbf{AR Scorpii} consisting of a red and a white dwarf exhibits highly variable emission. Episodic mass transfer between the dwarfs can be a reason for strong X-ray and radio emission. Further, 2 low mass single confirmed radio stars were found below MS. The reason for their deviation from MS is unknown.

\subsection{Planet hosting radio systems \& SPI}
The NASA exoplanet archive (NEA) catalog containing composite data for all the confirmed planets, was cross-matched with our sample to search for any 3~GHz planet-hosting radio source. We didn't find any match results. On cross-matching with TESS and Kepler candidates, we obtained a null result. However a K2 false positive, \textbf{EPIC 204165788.01} crossmatched with our sample. It turns out it is most likely an eclipsing binary companion  \citep{k2false} to the primary source \textbf{HIP 80474}.


\section{Sources Information Table}

A table containing the important cross-matching information for our entire sample of radio sources can be found in the machine readable format which is provided as a supplement to this article. An example table for our entire sample containing 20 rows with SIMBAD identifiers and object types has been provided below.

\begin{table*}
\movetabledown=5.2cm
\begin{rotatetable*}
\centering
\caption{}
\begin{tabular}{l l l l l l l l l l l}
\hline
\hline
\multicolumn{1}{c}{Gaia DR3 id} & \multicolumn{1}{c}{RAJ2016} & \multicolumn{1}{c}{DEJ2016} & \multicolumn{1}{c}{Gmag} & \multicolumn{1}{c}{BP-RP} & \multicolumn{1}{c}{Distance} & \multicolumn{1}{c}{$F_1$} & \multicolumn{1}{c}{$F_2$} & \multicolumn{1}{c}{SIMBAD id} & \multicolumn{1}{c}{Object type} & \multicolumn{1}{c}{Category} \\
\multicolumn{1}{c}{} & \multicolumn{1}{c}{(deg)} & \multicolumn{1}{c}{(deg)} & \multicolumn{1}{c}{} & \multicolumn{1}{c}{} & \multicolumn{1}{c}{(pc)} & \multicolumn{1}{c}{(mJy/beam)} & \multicolumn{1}{c}{(mJy/beam)} & \multicolumn{1}{c}{} & \multicolumn{1}{c}{} & \multicolumn{1}{c}{} \\
\hline
 2137474969751117312 & 295.961257 & 53.628387 & 15.56 & 2.77 & 97.60 & 1.02 & 1.05 & - & - & Confirmed \\
 6123452251070601216 & 215.462559 & -32.529713 & 17.43 & 3.01 & 129.21 & 1.84 & 1.50 & - & - & Confirmed \\
 3263936692671872512 & 54.196900 & 0.587042 & 5.61 & 1.24 & 29.43 & 14.42 & 29.89 & - & - & Confirmed \\
 852710613536664320 & 151.387126 & 54.869893 & 15.98 & 2.03 & 328.36 & 1.12 & 1.09 & - & - & Confirmed \\
 837388987723516160 & 162.460313 & 52.145178 & 14.59 & 1.34 & 333.83 & 0.86 & 0.97 & 2MASS & QSO\_Candidate & Confirmed \\
 & & & & & & & & J10495046+5208426& &\\
 3120983344792501760 & 98.146211 & 2.125139 & 17.71 & 3.08 & 219.15 & 9.49 & 9.33 & - & - & Confirmed \\
 4358897712402949632 & 246.359654 & -1.514962 & 19.38 & 3.02 & 271.37 & 3.34 & 2.77 & - & - & Confirmed \\
 3425077749587261440 & 92.366431 & 22.607449 & 9.48 & 0.98 & 110.87 & 1.10 & 1.20 & HD 252406 & RotV* & Confirmed \\
 2149628932626002432 & 271.589599 & 53.706930 & 19.15 & 3.33 & 219.77 & 1.30 & 1.32 & - & - & Confirmed \\
 2081589817375031552 & 303.704452 & 45.028526 & 7.43 & 1.29 & 242.18 & 4.92 & 12.33 & HD 192785 & SB* & Confirmed \\
 1847913321237991168 & 318.557201 & 27.130327 & 18.55 & 2.93 & 228.92 & 1.85 & 1.89 & - & - & Confirmed \\
 6237015550067391488 & 238.482701 & -23.978203 & 5.36 & 0 & 143.06 & 26.48 & 28.33 & PMN J1553-2358 & Radio & Confirmed \\
 1282848220976683264 & 220.644200 & 30.475568 & 13.66 & 1.42 & 213.28 & 0.84 & 0.94 & FIRST & Radio & Confirmed \\
 & & & & & & & & J144234.6+302832& &\\
 3314244361868724224 & 68.387760 & 18.016635 & 6.89 & 0.26 & 157.97 & 1.36 & 1.90 & HD 28867A & Star & Confirmed \\
 213484274324058624 & 81.250313 & 49.317184 & 10.47 & 1.32 & 273.44 & 3.00 & 2.11 & BD+49 1348 & RotV* & Confirmed \\
 861644351670829440 & 163.931091 & 60.469333 & 8.91 & 1.3 & 185.66 & 27.92 & 1.59 & V* DM UMa & RSCVnV* & Confirmed \\
 663853582908635520 & 126.572567 & 20.577690 & 15.35 & 1.81 & 309.44 & 2.45 & 2.84 & 2MASS & Star & Confirmed \\
 & & & & & & & & J08261745+2034402& &\\
 6235747125966268416 & 238.664642 & -25.243848 & 5.84 & -0.06 & 144.23 & 2.60 & 2.82 & * 3 Sco & RotV* & Confirmed \\
 2694353690543010304 & 324.417836 & 1.620191 & 12.16 & 2.97 & 35.91 & 1.38 & 1.83 & RX J2137.6+0137 & ** & Confirmed \\
 1005873614080407168 & 91.375153 & 60.819401 & 12.30 & 2.87 & 16.28 & 8.13 & 2.15 & - & - & Confirmed \\
 1328866562170960384 & 243.668787 & 33.858226 & 5.43 & 0.81 & 22.70 & 3.19 & 6.56 & - & - & Confirmed \\
 \hline
\end{tabular}
\hspace{2cm}
Table 1 is published in its entirety in the machine-readable format. A portion is shown here for guidance regarding its form and content.
\end{rotatetable*}
\end{table*}

\bibliography{sample631}{}

\begin{thebibliography}{}
\expandafter\ifx\csname natexlab\endcsname\relax\def\natexlab#1{#1}\fi
\providecommand{\url}[1]{\href{#1}{#1}}
\providecommand{\dodoi}[1]{doi:~\href{http://doi.org/#1}{\nolinkurl{#1}}}
\providecommand{\doeprint}[1]{\href{http://ascl.net/#1}{\nolinkurl{http://ascl.net/#1}}}
\providecommand{\doarXiv}[1]{\href{https://arxiv.org/abs/#1}{\nolinkurl{https://arxiv.org/abs/#1}}}

\bibitem[{Andersson {et~al.}(2022)Andersson, Fender, Lintott, Williams, Driessen, Woudt, van.der.Horst, Buckley, Motta, Rhodes, Eisner, Osten, Vreeswijk, Bloemen, \& Groot}]{meerkat1}
Andersson, A., Fender, R.~P., Lintott, C.~J., {et~al.} 2022, Monthly Notices of the Royal Astronomical Society, 513, 3482–3492, \dodoi{10.1093/mnras/stac1002}

\bibitem[{{Andrae} {et~al.}(2023){Andrae}, {Fouesneau}, {Sordo}, {Bailer-Jones}, {Dharmawardena}, {Rybizki}, {De Angeli}, {Lindstr{\o}m}, {Marshall}, {Drimmel}, {Korn}, {Soubiran}, {Brouillet}, {Casamiquela}, {Rix}, {Abreu Aramburu}, {{\'A}lvarez}, {Bakker}, {Bellas-Velidis}, {Bijaoui}, {Brugaletta}, {Burlacu}, {Carballo}, {Chaoul}, {Chiavassa}, {Contursi}, {Cooper}, {Creevey}, {Dafonte}, {Dapergolas}, {de Laverny}, {Delchambre}, {Demouchy}, {Edvardsson}, {Fr{\'e}mat}, {Garabato}, {Garc{\'\i}a-Lario}, {Garc{\'\i}a-Torres}, {Gavel}, {Gomez}, {Gonz{\'a}lez-Santamar{\'\i}a}, {Hatzidimitriou}, {Heiter}, {Jean-Antoine Piccolo}, {Kontizas}, {Kordopatis}, {Lanzafame}, {Lebreton}, {Licata}, {Livanou}, {Lobel}, {Lorca}, {Magdaleno Romeo}, {Manteiga}, {Marocco}, {Mary}, {Nicolas}, {Ordenovic}, {Pailler}, {Palicio}, {Pallas-Quintela}, {Panem}, {Pichon}, {Poggio}, {Recio-Blanco}, {Riclet}, {Robin}, {Santove{\~n}a}, {Sarro}, {Schultheis}, {Segol}, {Silvelo}, {Slezak}, {Smart}, {S{\"u}veges}, {Th{\'e}venin}, {Torralba
  Elipe}, {Ulla}, {Utrilla}, {Vallenari}, {van Dillen}, {Zhao}, \& {Zorec}}]{GSP_Phot}
{Andrae}, R., {Fouesneau}, M., {Sordo}, R., {et~al.} 2023, \aap, 674, A27, \dodoi{10.1051/0004-6361/202243462}

\bibitem[{Antonova {et~al.}(2013)Antonova, Hallinan, Doyle, Yu, Kuznetsov, Metodieva, Golden, \& Cruz}]{UCD}
Antonova, A., Hallinan, G., Doyle, J.~G., {et~al.} 2013, Astronomy \&amp; Astrophysics, 549, A131, \dodoi{10.1051/0004-6361/201118583}

\bibitem[{Banerjee {et~al.}(2024)Banerjee, Narang, Manoj, Henning, Tyagi, Surya, Nayak, \& Tripathi}]{bihan2024host}
Banerjee, B., Narang, M., Manoj, P., {et~al.} 2024, Host star properties of hot, warm and cold Jupiters in the solar neighborhood from \textit{Gaia} DR3: clues to formation pathways.
\newblock \doarXiv{2404.16499}

\bibitem[{Baraffe \& Chabrier(2018)}]{Baraffe_2018}
Baraffe, I., \& Chabrier, G. 2018, Astronomy \&amp; Astrophysics, 619, A177, \dodoi{10.1051/0004-6361/201834062}

\bibitem[{{Becker} {et~al.}(1995){Becker}, {White}, \& {Helfand}}]{first}
{Becker}, R.~H., {White}, R.~L., \& {Helfand}, D.~J. 1995, \apj, 450, 559, \dodoi{10.1086/176166}

\bibitem[{{Benz} \& {G{\"u}del}(2010)}]{gudel&benz}
{Benz}, A.~O., \& {G{\"u}del}, M. 2010, \araa, 48, 241, \dodoi{10.1146/annurev-astro-082708-101757}

\bibitem[{{Benz} \& {Guedel}(1994)}]{benz_gudel}
{Benz}, A.~O., \& {Guedel}, M. 1994, \aap, 285, 621

\bibitem[{Berger(2002)}]{Berger_2002}
Berger, E. 2002, The Astrophysical Journal, 572, 503–513, \dodoi{10.1086/340301}

\bibitem[{Best(2020)}]{ucd_sheet}
Best, W. M.~J. 2020, The UltracoolSheet: Photometry, Astrometry, Spectroscopy, and Multiplicity for 3000+ Ultracool Dwarfs and Imaged Exoplanets, \dodoi{10.5281/zenodo.4570814}

\bibitem[{{Bieging} {et~al.}(1989){Bieging}, {Abbott}, \& {Churchwell}}]{OBsurvey}
{Bieging}, J.~H., {Abbott}, D.~C., \& {Churchwell}, E.~B. 1989, \apj, 340, 518, \dodoi{10.1086/167414}

\bibitem[{Bookbinder(1988)}]{bookbinder}
Bookbinder, A.~J. 1988, Astrophysics and Space Science Library, 143.
\newblock \url{https://doi.org/10.1007/978-94-009-2951-7_42}

\bibitem[{Brown {et~al.}(2018)Brown, Vallenari, Prusti, de~Bruijne, Babusiaux, Bailer-Jones, Biermann, Evans, Eyer, Jansen, Jordi, Klioner, Lammers, Lindegren, Luri, Mignard, Panem, Pourbaix, Randich, Sartoretti, Siddiqui, Soubiran, van Leeuwen, Walton, Arenou, Bastian, Cropper, Drimmel, Katz, Lattanzi, Bakker, Cacciari, Castañeda, Chaoul, Cheek, De~Angeli, Fabricius, Guerra, Holl, Masana, Messineo, Mowlavi, Nienartowicz, Panuzzo, Portell, Riello, Seabroke, Tanga, Thévenin, Gracia-Abril, Comoretto, Garcia-Reinaldos, Teyssier, Altmann, Andrae, Audard, Bellas-Velidis, Benson, Berthier, Blomme, Burgess, Busso, Carry, Cellino, Clementini, Clotet, Creevey, Davidson, De~Ridder, Delchambre, Dell’Oro, Ducourant, Fernández-Hernández, Fouesneau, Frémat, Galluccio, García-Torres, González-Núñez, González-Vidal, Gosset, Guy, Halbwachs, Hambly, Harrison, Hernández, Hestroffer, Hodgkin, Hutton, Jasniewicz, Jean-Antoine-Piccolo, Jordan, Korn, Krone-Martins, Lanzafame, Lebzelter, Löffler, Manteiga, Marrese,
  Martín-Fleitas, Moitinho, Mora, Muinonen, Osinde, Pancino, Pauwels, Petit, Recio-Blanco, Richards, Rimoldini, Robin, Sarro, Siopis, Smith, Sozzetti, Süveges, Torra, van Reeven, Abbas, Abreu~Aramburu, Accart, Aerts, Altavilla, Álvarez, Alvarez, Alves, Anderson, Andrei, Anglada~Varela, Antiche, Antoja, Arcay, Astraatmadja, Bach, Baker, Balaguer-Núñez, Balm, Barache, Barata, Barbato, Barblan, Barklem, Barrado, Barros, Barstow, Bartholomé~Muñoz, Bassilana, Becciani, Bellazzini, Berihuete, Bertone, Bianchi, Bienaymé, Blanco-Cuaresma, Boch, Boeche, Bombrun, Borrachero, Bossini, Bouquillon, Bourda, Bragaglia, Bramante, Breddels, Bressan, Brouillet, Brüsemeister, Brugaletta, Bucciarelli, Burlacu, Busonero, Butkevich, Buzzi, Caffau, Cancelliere, Cannizzaro, Cantat-Gaudin, Carballo, Carlucci, Carrasco, Casamiquela, Castellani, Castro-Ginard, Charlot, Chemin, Chiavassa, Cocozza, Costigan, Cowell, Crifo, Crosta, Crowley, Cuypers†, Dafonte, Damerdji, Dapergolas, David, David, de~Laverny, De~Luise, De~March,
  de~Martino, de~Souza, de~Torres, Debosscher, del Pozo, Delbo, Delgado, Delgado, Di~Matteo, Diakite, Diener, Distefano, Dolding, Drazinos, Durán, Edvardsson, Enke, Eriksson, Esquej, Eynard~Bontemps, Fabre, Fabrizio, Faigler, Falcão, Farràs~Casas, Federici, Fedorets, Fernique, Figueras, Filippi, Findeisen, Fonti, Fraile, Fraser, Frézouls, Gai, Galleti, Garabato, García-Sedano, Garofalo, Garralda, Gavel, Gavras, Gerssen, Geyer, Giacobbe, Gilmore, Girona, Giuffrida, Glass, Gomes, Granvik, Gueguen, Guerrier, Guiraud, Gutiérrez-Sánchez, Haigron, Hatzidimitriou, Hauser, Haywood, Heiter, Helmi, Heu, Hilger, Hobbs, Hofmann, Holland, Huckle, Hypki, Icardi, Janßen, Jevardat~de Fombelle, Jonker, Juhász, Julbe, Karampelas, Kewley, Klar, Kochoska, Kohley, Kolenberg, Kontizas, Kontizas, Koposov, Kordopatis, Kostrzewa-Rutkowska, Koubsky, Lambert, Lanza, Lasne, Lavigne, Le~Fustec, Le~Poncin-Lafitte, Lebreton, Leccia, Leclerc, Lecoeur-Taibi, Lenhardt, Leroux, Liao, Licata, Lindstrøm, Lister, Livanou, Lobel, López,
  Managau, Mann, Mantelet, Marchal, Marchant, Marconi, Marinoni, Marschalkó, Marshall, Martino, Marton, Mary, Massari, Matijevič, Mazeh, McMillan, Messina, Michalik, Millar, Molina, Molinaro, Molnár, Montegriffo, Mor, Morbidelli, Morel, Morris, Mulone, Muraveva, Musella, Nelemans, Nicastro, Noval, O’Mullane, Ordénovic, Ordóñez-Blanco, Osborne, Pagani, Pagano, Pailler, Palacin, Palaversa, Panahi, Pawlak, Piersimoni, Pineau, Plachy, Plum, Poggio, Poujoulet, Prša, Pulone, Racero, Ragaini, Rambaux, Ramos-Lerate, Regibo, Reylé, Riclet, Ripepi, Riva, Rivard, Rixon, Roegiers, Roelens, Romero-Gómez, Rowell, Royer, Ruiz-Dern, Sadowski, Sagristà~Sellés, Sahlmann, Salgado, Salguero, Sanna, Santana-Ros, Sarasso, Savietto, Schultheis, Sciacca, Segol, Segovia, Ségransan, Shih, Siltala, Silva, Smart, Smith, Solano, Solitro, Sordo, Soria~Nieto, Souchay, Spagna, Spoto, Stampa, Steele, Steidelmüller, Stephenson, Stoev, Suess, Surdej, Szabados, Szegedi-Elek, Tapiador, Taris, Tauran, Taylor, Teixeira, Terrett,
  Teyssandier, Thuillot, Titarenko, Torra~Clotet, Turon, Ulla, Utrilla, Uzzi, Vaillant, Valentini, Valette, van Elteren, Van~Hemelryck, van Leeuwen, Vaschetto, Vecchiato, Veljanoski, Viala, Vicente, Vogt, von Essen, Voss, Votruba, Voutsinas, Walmsley, Weiler, Wertz, Wevers, Wyrzykowski, Yoldas, Žerjal, Ziaeepour, Zorec, Zschocke, Zucker, Zurbach, \& Zwitter}]{gaiadr2}
Brown, A. G.~A., Vallenari, A., Prusti, T., {et~al.} 2018, Astronomy \&amp; Astrophysics, 616, A1, \dodoi{10.1051/0004-6361/201833051}

\bibitem[{{Burningham} {et~al.}(2016){Burningham}, {Hardcastle}, {Nichols}, {Casewell}, {Littlefair}, {Stark}, {Burleigh}, {Metchev}, {Tannock}, {van Weeren}, {Williams}, \& {Wynn}}]{Burningham}
{Burningham}, B., {Hardcastle}, M., {Nichols}, J.~D., {et~al.} 2016, \mnras, 463, 2202, \dodoi{10.1093/mnras/stw2065}

\bibitem[{Callingham {et~al.}(2021)Callingham, Vedantham, Shimwell, Pope, Davis, Best, Hardcastle, Röttgering, Sabater, Tasse, van Weeren, Williams, Zarka, de~Gasperin, \& Drabent}]{calingham}
Callingham, J.~R., Vedantham, H.~K., Shimwell, T.~W., {et~al.} 2021, Nature Astronomy, 5, 1233–1239, \dodoi{10.1038/s41550-021-01483-0}

\bibitem[{Cauley {et~al.}(2019)Cauley, Shkolnik, Llama, \& Lanza}]{Cauley+19}
Cauley, P.~W., Shkolnik, E.~L., Llama, J., \& Lanza, A.~F. 2019, Nature Astronomy, 3, 1128–1134, \dodoi{10.1038/s41550-019-0840-x}

\bibitem[{Ceballos {et~al.}(2024)Ceballos, Cendes, Berger, \& Williams}]{vol_lim_exo24}
Ceballos, K. N.~O., Cendes, Y., Berger, E., \& Williams, P. K.~G. 2024, A Volume-Limited Radio Search for Magnetic Activity in 140 Exoplanets with the Very Large Array.
\newblock \doarXiv{2404.16940}

\bibitem[{Chabrier \& Baraffe(2000)}]{dwarfs2}
Chabrier, G., \& Baraffe, I. 2000, Annual Review of Astronomy and Astrophysics, 38, 337–377, \dodoi{10.1146/annurev.astro.38.1.337}

\bibitem[{{Condon} {et~al.}(1998){Condon}, {Cotton}, {Greisen}, {Yin}, {Perley}, {Taylor}, \& {Broderick}}]{nvss}
{Condon}, J.~J., {Cotton}, W.~D., {Greisen}, E.~W., {et~al.} 1998, \aj, 115, 1693, \dodoi{10.1086/300337}

\bibitem[{{Das} {et~al.}(2020){Das}, {Chandra}, {Wade}, {Shultz}, \& {Sikora}}]{Das_2020}
{Das}, B., {Chandra}, P., {Wade}, G.~A., {Shultz}, M.~E., \& {Sikora}, J. 2020, in Stellar Magnetism: A Workshop in Honour of the Career and Contributions of John D. Landstreet, ed. G.~{Wade}, E.~{Alecian}, D.~{Bohlender}, \& A.~{Sigut}, Vol.~11, 66--73, \dodoi{10.48550/arXiv.1912.09430}

\bibitem[{Das {et~al.}(2022)Das, Chandra, Shultz, Wade, Sikora, Kochukhov, Neiner, Oksala, \& Alecian}]{Das_2022}
Das, B., Chandra, P., Shultz, M.~E., {et~al.} 2022, The Astrophysical Journal, 925, 125, \dodoi{10.3847/1538-4357/ac2576}

\bibitem[{Davis {et~al.}(2021)Davis, Vedantham, Callingham, Shimwell, Vidotto, Zarka, Ray, \& Drabent}]{wxuma}
Davis, I., Vedantham, H.~K., Callingham, J.~R., {et~al.} 2021, Astronomy \&amp; Astrophysics, 650, L20, \dodoi{10.1051/0004-6361/202140772}

\bibitem[{{Donati} \& {Landstreet}(2009)}]{star_mag}
{Donati}, J.~F., \& {Landstreet}, J.~D. 2009, \araa, 47, 333, \dodoi{10.1146/annurev-astro-082708-101833}

\bibitem[{{Dorman} {et~al.}(1989){Dorman}, {Nelson}, \& {Chau}}]{dwarfs1}
{Dorman}, B., {Nelson}, L.~A., \& {Chau}, W.~Y. 1989, \apj, 342, 1003, \dodoi{10.1086/167658}

\bibitem[{{Doyle} {et~al.}(2019){Doyle}, {Ramsay}, {Doyle}, \& {Wu}}]{Erup*}
{Doyle}, L., {Ramsay}, G., {Doyle}, J.~G., \& {Wu}, K. 2019, \mnras, 489, 437, \dodoi{10.1093/mnras/stz2205}

\bibitem[{Driessen {et~al.}(2023)Driessen, Heald, Duchesne, Murphy, Lenc, Leung, \& Moss}]{pm_method}
Driessen, L.~N., Heald, G., Duchesne, S.~W., {et~al.} 2023, Publications of the Astronomical Society of Australia, 40, \dodoi{10.1017/pasa.2023.26}

\bibitem[{Driessen {et~al.}(2021)Driessen, Williams, McDonald, Stappers, Buckley, Fender, \& Woudt}]{meerkat2}
Driessen, L.~N., Williams, D. R.~A., McDonald, I., {et~al.} 2021, Monthly Notices of the Royal Astronomical Society, 510, 1083–1092, \dodoi{10.1093/mnras/stab3461}

\bibitem[{{Driessen} {et~al.}(2024){Driessen}, {Pritchard}, {Murphy}, {Heald}, {Robrade}, {Das}, {Duchesne}, {Kaplan}, {Lenc}, {Lynch}, {Pope}, {Rose}, {Stelzer}, {Wang}, \& {Zic}}]{Sydney}
{Driessen}, L.~N., {Pritchard}, J., {Murphy}, T., {et~al.} 2024, arXiv e-prints, arXiv:2404.07418, \dodoi{10.48550/arXiv.2404.07418}

\bibitem[{{Dulk}(1985)}]{dulk}
{Dulk}, G.~A. 1985, \araa, 23, 169, \dodoi{10.1146/annurev.aa.23.090185.001125}

\bibitem[{Feeney-Johansson {et~al.}(2021)Feeney-Johansson, Purser, Ray, Vidotto, Eislöffel, Callingham, Shimwell, Vedantham, Hallinan, \& Tasse}]{WTTS}
Feeney-Johansson, A., Purser, S. J.~D., Ray, T.~P., {et~al.} 2021, Astronomy \&amp; Astrophysics, 653, A101, \dodoi{10.1051/0004-6361/202140849}

\bibitem[{{Frasca} {et~al.}(2018){Frasca}, {Guillout}, {Klutsch}, {Ferrero}, {Marilli}, {Biazzo}, {Gandolfi}, \& {Montes}}]{SB*}
{Frasca}, A., {Guillout}, P., {Klutsch}, A., {et~al.} 2018, \aap, 612, A96, \dodoi{10.1051/0004-6361/201732028}

\bibitem[{{Gaia Collaboration} {et~al.}(2021){Gaia Collaboration}, {Smart}, {Sarro}, {Rybizki}, {Reyl{\'e}}, {Robin}, {Hambly}, {Abbas}, {Barstow}, {de Bruijne}, {Bucciarelli}, {Carrasco}, {Cooper}, {Hodgkin}, {Masana}, {Michalik}, {Sahlmann}, {Sozzetti}, {Brown}, {Vallenari}, {Prusti}, {Babusiaux}, {Biermann}, {Creevey}, {Evans}, {Eyer}, {Hutton}, {Jansen}, {Jordi}, {Klioner}, {Lammers}, {Lindegren}, {Luri}, {Mignard}, {Panem}, {Pourbaix}, {Randich}, {Sartoretti}, {Soubiran}, {Walton}, {Arenou}, {Bailer-Jones}, {Bastian}, {Cropper}, {Drimmel}, {Katz}, {Lattanzi}, {van Leeuwen}, {Bakker}, {Casta{\~n}eda}, {De Angeli}, {Ducourant}, {Fabricius}, {Fouesneau}, {Fr{\'e}mat}, {Guerra}, {Guerrier}, {Guiraud}, {Jean-Antoine Piccolo}, {Messineo}, {Mowlavi}, {Nicolas}, {Nienartowicz}, {Pailler}, {Panuzzo}, {Riclet}, {Roux}, {Seabroke}, {Sordo}, {Tanga}, {Th{\'e}venin}, {Gracia-Abril}, {Portell}, {Teyssier}, {Altmann}, {Andrae}, {Bellas-Velidis}, {Benson}, {Berthier}, {Blomme}, {Brugaletta}, {Burgess}, {Busso}, {Carry},
  {Cellino}, {Cheek}, {Clementini}, {Damerdji}, {Davidson}, {Delchambre}, {Dell'Oro}, {Fern{\'a}ndez-Hern{\'a}ndez}, {Galluccio}, {Garc{\'\i}a-Lario}, {Garcia-Reinaldos}, {Gonz{\'a}lez-N{\'u}{\~n}ez}, {Gosset}, {Haigron}, {Halbwachs}, {Harrison}, {Hatzidimitriou}, {Heiter}, {Hern{\'a}ndez}, {Hestroffer}, {Holl}, {Jan{\ss}en}, {Jevardat de Fombelle}, {Jordan}, {Krone-Martins}, {Lanzafame}, {L{\"o}ffler}, {Lorca}, {Manteiga}, {Marchal}, {Marrese}, {Moitinho}, {Mora}, {Muinonen}, {Osborne}, {Pancino}, {Pauwels}, {Recio-Blanco}, {Richards}, {Riello}, {Rimoldini}, {Roegiers}, {Siopis}, {Smith}, {Ulla}, {Utrilla}, {van Leeuwen}, {van Reeven}, {Abreu Aramburu}, {Accart}, {Aerts}, {Aguado}, {Ajaj}, {Altavilla}, {{\'A}lvarez}, {{\'A}lvarez Cid-Fuentes}, {Alves}, {Anderson}, {Anglada Varela}, {Antoja}, {Audard}, {Baines}, {Baker}, {Balaguer-N{\'u}{\~n}ez}, {Balbinot}, {Balog}, {Barache}, {Barbato}, {Barros}, {Bartolom{\'e}}, {Bassilana}, {Bauchet}, {Baudesson-Stella}, {Becciani}, {Bellazzini}, {Bernet}, {Bertone},
  {Bianchi}, {Blanco-Cuaresma}, {Boch}, {Bombrun}, {Bossini}, {Bouquillon}, {Bragaglia}, {Bramante}, {Breedt}, {Bressan}, {Brouillet}, {Burlacu}, {Busonero}, {Butkevich}, {Buzzi}, {Caffau}, {Cancelliere}, {C{\'a}novas}, {Cantat-Gaudin}, {Carballo}, {Carlucci}, {Carnerero}, {Casamiquela}, {Castellani}, {Castro-Ginard}, {Castro Sampol}, {Chaoul}, {Charlot}, {Chemin}, {Chiavassa}, {Cioni}, {Comoretto}, {Cornez}, {Cowell}, {Crifo}, {Crosta}, {Crowley}, {Dafonte}, {Dapergolas}, {David}, {David}, {de Laverny}, {De Luise}, {De March}, {De Ridder}, {de Souza}, {de Teodoro}, {de Torres}, {del Peloso}, {del Pozo}, {Delgado}, {Delgado}, {Delisle}, {Di Matteo}, {Diakite}, {Diener}, {Distefano}, {Dolding}, {Eappachen}, {Edvardsson}, {Enke}, {Esquej}, {Fabre}, {Fabrizio}, {Faigler}, {Fedorets}, {Fernique}, {Fienga}, {Figueras}, {Fouron}, {Fragkoudi}, {Fraile}, {Franke}, {Gai}, {Garabato}, {Garcia-Gutierrez}, {Garc{\'\i}a-Torres}, {Garofalo}, {Gavras}, {Gerlach}, {Geyer}, {Giacobbe}, {Gilmore}, {Girona}, {Giuffrida},
  {Gomel}, {Gomez}, {Gonzalez-Santamaria}, {Gonz{\'a}lez-Vidal}, {Granvik}, {Guti{\'e}rrez-S{\'a}nchez}, {Guy}, {Hauser}, {Haywood}, {Helmi}, {Hidalgo}, {Hilger}, {H{\l}adczuk}, {Hobbs}, {Holland}, {Huckle}, {Jasniewicz}, {Jonker}, {Juaristi Campillo}, {Julbe}, {Karbevska}, {Kervella}, {Khanna}, {Kochoska}, {Kontizas}, {Kordopatis}, {Korn}, {Kostrzewa-Rutkowska}, {Kruszy{\'n}ska}, {Lambert}, {Lanza}, {Lasne}, {Le Campion}, {Le Fustec}, {Lebreton}, {Lebzelter}, {Leccia}, {Leclerc}, {Lecoeur-Taibi}, {Liao}, {Licata}, {Lindstr{\o}m}, {Lister}, {Livanou}, {Lobel}, {Madrero Pardo}, {Managau}, {Mann}, {Marchant}, {Marconi}, {Marcos Santos}, {Marinoni}, {Marocco}, {Marshall}, {Martin Polo}, {Mart{\'\i}n-Fleitas}, {Masip}, {Massari}, {Mastrobuono-Battisti}, {Mazeh}, {McMillan}, {Messina}, {Millar}, {Mints}, {Molina}, {Molinaro}, {Moln{\'a}r}, {Montegriffo}, {Mor}, {Morbidelli}, {Morel}, {Morris}, {Mulone}, {Munoz}, {Muraveva}, {Murphy}, {Musella}, {Noval}, {Ord{\'e}novic}, {Orr{\`u}}, {Osinde}, {Pagani}, {Pagano},
  {Palaversa}, {Palicio}, {Panahi}, {Pawlak}, {Pe{\~n}alosa Esteller}, {Penttil{\"a}}, {Piersimoni}, {Pineau}, {Plachy}, {Plum}, {Poggio}, {Poretti}, {Poujoulet}, {Pr{\v{s}}a}, {Pulone}, {Racero}, {Ragaini}, {Rainer}, {Raiteri}, {Rambaux}, {Ramos}, {Ramos-Lerate}, {Re Fiorentin}, {Regibo}, {Ripepi}, {Riva}, {Rixon}, {Robichon}, {Robin}, {Roelens}, {Rohrbasser}, {Romero-G{\'o}mez}, {Rowell}, {Royer}, {Rybicki}, {Sadowski}, {Sagrist{\`a} Sell{\'e}s}, {Salgado}, {Salguero}, {Samaras}, {Sanchez Gimenez}, {Sanna}, {Santove{\~n}a}, {Sarasso}, {Schultheis}, {Sciacca}, {Segol}, {Segovia}, {S{\'e}gransan}, {Semeux}, {Shahaf}, {Siddiqui}, {Siebert}, {Siltala}, {Slezak}, {Solano}, {Solitro}, {Souami}, {Souchay}, {Spagna}, {Spoto}, {Steele}, {Steidelm{\"u}ller}, {Stephenson}, {S{\"u}veges}, {Szabados}, {Szegedi-Elek}, {Taris}, {Tauran}, {Taylor}, {Teixeira}, {Thuillot}, {Tonello}, {Torra}, {Torra}, {Turon}, {Unger}, {Vaillant}, {van Dillen}, {Vanel}, {Vecchiato}, {Viala}, {Vicente}, {Voutsinas}, {Weiler}, {Wevers},
  {Wyrzykowski}, {Yoldas}, {Yvard}, {Zhao}, {Zorec}, {Zucker}, {Zurbach}, \& {Zwitter}}]{GCNS}
{Gaia Collaboration}, {Smart}, R.~L., {Sarro}, L.~M., {et~al.} 2021, \aap, 649, A6, \dodoi{10.1051/0004-6361/202039498}

\bibitem[{{Gaia Collaboration} {et~al.}(2023){Gaia Collaboration}, {Vallenari}, {Brown}, {Prusti}, {de Bruijne}, {Arenou}, {Babusiaux}, {Biermann}, {Creevey}, {Ducourant}, {Evans}, {Eyer}, {Guerra}, {Hutton}, {Jordi}, {Klioner}, {Lammers}, {Lindegren}, {Luri}, {Mignard}, {Panem}, {Pourbaix}, {Randich}, {Sartoretti}, {Soubiran}, {Tanga}, {Walton}, {Bailer-Jones}, {Bastian}, {Drimmel}, {Jansen}, {Katz}, {Lattanzi}, {van Leeuwen}, {Bakker}, {Cacciari}, {Casta{\~n}eda}, {De Angeli}, {Fabricius}, {Fouesneau}, {Fr{\'e}mat}, {Galluccio}, {Guerrier}, {Heiter}, {Masana}, {Messineo}, {Mowlavi}, {Nicolas}, {Nienartowicz}, {Pailler}, {Panuzzo}, {Riclet}, {Roux}, {Seabroke}, {Sordo}, {Th{\'e}venin}, {Gracia-Abril}, {Portell}, {Teyssier}, {Altmann}, {Andrae}, {Audard}, {Bellas-Velidis}, {Benson}, {Berthier}, {Blomme}, {Burgess}, {Busonero}, {Busso}, {C{\'a}novas}, {Carry}, {Cellino}, {Cheek}, {Clementini}, {Damerdji}, {Davidson}, {de Teodoro}, {Nu{\~n}ez Campos}, {Delchambre}, {Dell'Oro}, {Esquej},
  {Fern{\'a}ndez-Hern{\'a}ndez}, {Fraile}, {Garabato}, {Garc{\'\i}a-Lario}, {Gosset}, {Haigron}, {Halbwachs}, {Hambly}, {Harrison}, {Hern{\'a}ndez}, {Hestroffer}, {Hodgkin}, {Holl}, {Jan{\ss}en}, {Jevardat de Fombelle}, {Jordan}, {Krone-Martins}, {Lanzafame}, {L{\"o}ffler}, {Marchal}, {Marrese}, {Moitinho}, {Muinonen}, {Osborne}, {Pancino}, {Pauwels}, {Recio-Blanco}, {Reyl{\'e}}, {Riello}, {Rimoldini}, {Roegiers}, {Rybizki}, {Sarro}, {Siopis}, {Smith}, {Sozzetti}, {Utrilla}, {van Leeuwen}, {Abbas}, {{\'A}brah{\'a}m}, {Abreu Aramburu}, {Aerts}, {Aguado}, {Ajaj}, {Aldea-Montero}, {Altavilla}, {{\'A}lvarez}, {Alves}, {Anders}, {Anderson}, {Anglada Varela}, {Antoja}, {Baines}, {Baker}, {Balaguer-N{\'u}{\~n}ez}, {Balbinot}, {Balog}, {Barache}, {Barbato}, {Barros}, {Barstow}, {Bartolom{\'e}}, {Bassilana}, {Bauchet}, {Becciani}, {Bellazzini}, {Berihuete}, {Bernet}, {Bertone}, {Bianchi}, {Binnenfeld}, {Blanco-Cuaresma}, {Blazere}, {Boch}, {Bombrun}, {Bossini}, {Bouquillon}, {Bragaglia}, {Bramante}, {Breedt},
  {Bressan}, {Brouillet}, {Brugaletta}, {Bucciarelli}, {Burlacu}, {Butkevich}, {Buzzi}, {Caffau}, {Cancelliere}, {Cantat-Gaudin}, {Carballo}, {Carlucci}, {Carnerero}, {Carrasco}, {Casamiquela}, {Castellani}, {Castro-Ginard}, {Chaoul}, {Charlot}, {Chemin}, {Chiaramida}, {Chiavassa}, {Chornay}, {Comoretto}, {Contursi}, {Cooper}, {Cornez}, {Cowell}, {Crifo}, {Cropper}, {Crosta}, {Crowley}, {Dafonte}, {Dapergolas}, {David}, {David}, {de Laverny}, {De Luise}, {De March}, {De Ridder}, {de Souza}, {de Torres}, {del Peloso}, {del Pozo}, {Delbo}, {Delgado}, {Delisle}, {Demouchy}, {Dharmawardena}, {Di Matteo}, {Diakite}, {Diener}, {Distefano}, {Dolding}, {Edvardsson}, {Enke}, {Fabre}, {Fabrizio}, {Faigler}, {Fedorets}, {Fernique}, {Fienga}, {Figueras}, {Fournier}, {Fouron}, {Fragkoudi}, {Gai}, {Garcia-Gutierrez}, {Garcia-Reinaldos}, {Garc{\'\i}a-Torres}, {Garofalo}, {Gavel}, {Gavras}, {Gerlach}, {Geyer}, {Giacobbe}, {Gilmore}, {Girona}, {Giuffrida}, {Gomel}, {Gomez}, {Gonz{\'a}lez-N{\'u}{\~n}ez},
  {Gonz{\'a}lez-Santamar{\'\i}a}, {Gonz{\'a}lez-Vidal}, {Granvik}, {Guillout}, {Guiraud}, {Guti{\'e}rrez-S{\'a}nchez}, {Guy}, {Hatzidimitriou}, {Hauser}, {Haywood}, {Helmer}, {Helmi}, {Sarmiento}, {Hidalgo}, {Hilger}, {H{\l}adczuk}, {Hobbs}, {Holland}, {Huckle}, {Jardine}, {Jasniewicz}, {Jean-Antoine Piccolo}, {Jim{\'e}nez-Arranz}, {Jorissen}, {Juaristi Campillo}, {Julbe}, {Karbevska}, {Kervella}, {Khanna}, {Kontizas}, {Kordopatis}, {Korn}, {K{\'o}sp{\'a}l}, {Kostrzewa-Rutkowska}, {Kruszy{\'n}ska}, {Kun}, {Laizeau}, {Lambert}, {Lanza}, {Lasne}, {Le Campion}, {Lebreton}, {Lebzelter}, {Leccia}, {Leclerc}, {Lecoeur-Taibi}, {Liao}, {Licata}, {Lindstr{\o}m}, {Lister}, {Livanou}, {Lobel}, {Lorca}, {Loup}, {Madrero Pardo}, {Magdaleno Romeo}, {Managau}, {Mann}, {Manteiga}, {Marchant}, {Marconi}, {Marcos}, {Marcos Santos}, {Mar{\'\i}n Pina}, {Marinoni}, {Marocco}, {Marshall}, {Martin Polo}, {Mart{\'\i}n-Fleitas}, {Marton}, {Mary}, {Masip}, {Massari}, {Mastrobuono-Battisti}, {Mazeh}, {McMillan}, {Messina}, {Michalik},
  {Millar}, {Mints}, {Molina}, {Molinaro}, {Moln{\'a}r}, {Monari}, {Mongui{\'o}}, {Montegriffo}, {Montero}, {Mor}, {Mora}, {Morbidelli}, {Morel}, {Morris}, {Muraveva}, {Murphy}, {Musella}, {Nagy}, {Noval}, {Oca{\~n}a}, {Ogden}, {Ordenovic}, {Osinde}, {Pagani}, {Pagano}, {Palaversa}, {Palicio}, {Pallas-Quintela}, {Panahi}, {Payne-Wardenaar}, {Pe{\~n}alosa Esteller}, {Penttil{\"a}}, {Pichon}, {Piersimoni}, {Pineau}, {Plachy}, {Plum}, {Poggio}, {Pr{\v{s}}a}, {Pulone}, {Racero}, {Ragaini}, {Rainer}, {Raiteri}, {Rambaux}, {Ramos}, {Ramos-Lerate}, {Re Fiorentin}, {Regibo}, {Richards}, {Rios Diaz}, {Ripepi}, {Riva}, {Rix}, {Rixon}, {Robichon}, {Robin}, {Robin}, {Roelens}, {Rogues}, {Rohrbasser}, {Romero-G{\'o}mez}, {Rowell}, {Royer}, {Ruz Mieres}, {Rybicki}, {Sadowski}, {S{\'a}ez N{\'u}{\~n}ez}, {Sagrist{\`a} Sell{\'e}s}, {Sahlmann}, {Salguero}, {Samaras}, {Sanchez Gimenez}, {Sanna}, {Santove{\~n}a}, {Sarasso}, {Schultheis}, {Sciacca}, {Segol}, {Segovia}, {S{\'e}gransan}, {Semeux}, {Shahaf}, {Siddiqui}, {Siebert},
  {Siltala}, {Silvelo}, {Slezak}, {Slezak}, {Smart}, {Snaith}, {Solano}, {Solitro}, {Souami}, {Souchay}, {Spagna}, {Spina}, {Spoto}, {Steele}, {Steidelm{\"u}ller}, {Stephenson}, {S{\"u}veges}, {Surdej}, {Szabados}, {Szegedi-Elek}, {Taris}, {Taylor}, {Teixeira}, {Tolomei}, {Tonello}, {Torra}, {Torra}, {Torralba Elipe}, {Trabucchi}, {Tsounis}, {Turon}, {Ulla}, {Unger}, {Vaillant}, {van Dillen}, {van Reeven}, {Vanel}, {Vecchiato}, {Viala}, {Vicente}, {Voutsinas}, {Weiler}, {Wevers}, {Wyrzykowski}, {Yoldas}, {Yvard}, {Zhao}, {Zorec}, {Zucker}, \& {Zwitter}}]{gaia_dr3}
{Gaia Collaboration}, {Vallenari}, A., {Brown}, A.~G.~A., {et~al.} 2023, \aap, 674, A1, \dodoi{10.1051/0004-6361/202243940}

\bibitem[{Galli {et~al.}(2018)Galli, Loinard, Ortiz-Léon, Kounkel, Dzib, Mioduszewski, Rodríguez, Hartmann, Teixeira, Torres, Rivera, Boden, Evans~II, Briceño, Tobin, \& Heyer}]{vlbi_tau}
Galli, P. A.~B., Loinard, L., Ortiz-Léon, G.~N., {et~al.} 2018, The Astrophysical Journal, 859, 33, \dodoi{10.3847/1538-4357/aabf91}

\bibitem[{{Goldreich} \& {Lynden-Bell}(1969)}]{Jupiter-Io}
{Goldreich}, P., \& {Lynden-Bell}, D. 1969, \apj, 156, 59, \dodoi{10.1086/149947}

\bibitem[{{Gordon} {et~al.}(2020){Gordon}, {Boyce}, {O'Dea}, {Rudnick}, {Andernach}, {Vantyghem}, {Baum}, {Bui}, \& {Dionyssiou}}]{vlassql1}
{Gordon}, Y.~A., {Boyce}, M.~M., {O'Dea}, C.~P., {et~al.} 2020, Research Notes of the American Astronomical Society, 4, 175, \dodoi{10.3847/2515-5172/abbe23}

\bibitem[{{Grie{\ss}meier}(2015)}]{spi}
{Grie{\ss}meier}, J.-M. 2015, in Astrophysics and Space Science Library, Vol. 411, Characterizing Stellar and Exoplanetary Environments, ed. H.~{Lammer} \& M.~{Khodachenko}, 213, \dodoi{10.1007/978-3-319-09749-7_11}

\bibitem[{Gudel(2002)}]{gudel_review}
Gudel, M. 2002, Annual Review of Astronomy and Astrophysics, 40, 217–261, \dodoi{10.1146/annurev.astro.40.060401.093806}

\bibitem[{{Gudel} {et~al.}(1993){Gudel}, {Schmitt}, {Bookbinder}, \& {Fleming}}]{gudel93}
{Gudel}, M., {Schmitt}, J. H.~M.~M., {Bookbinder}, J.~A., \& {Fleming}, T.~A. 1993, \apj, 415, 236, \dodoi{10.1086/173158}

\bibitem[{{Intema} {et~al.}(2017){Intema}, {Jagannathan}, {Mooley}, \& {Frail}}]{tgss}
{Intema}, H.~T., {Jagannathan}, P., {Mooley}, K.~P., \& {Frail}, D.~A. 2017, \aap, 598, A78, \dodoi{10.1051/0004-6361/201628536}

\bibitem[{Jao {et~al.}(2018)Jao, Henry, Gies, \& Hambly}]{Jao_2018}
Jao, W.-C., Henry, T.~J., Gies, D.~R., \& Hambly, N.~C. 2018, The Astrophysical Journal Letters, 861, L11, \dodoi{10.3847/2041-8213/aacdf6}

\bibitem[{{Kao} {et~al.}(2017){Kao}, {Hallinan}, {Pineda}, {Escala}, {Burgasser}, \& {Stevenson}}]{Kao17}
{Kao}, M., {Hallinan}, G., {Pineda}, J.~S., {et~al.} 2017, in American Astronomical Society Meeting Abstracts, Vol. 229, American Astronomical Society Meeting Abstracts \#229, 408.06

\bibitem[{Kavanagh {et~al.}(2021)Kavanagh, Vidotto, Klein, Jardine, Donati, \& Ó Fionnagáin}]{Kavanagh_spi}
Kavanagh, R.~D., Vidotto, A.~A., Klein, B., {et~al.} 2021, Monthly Notices of the Royal Astronomical Society, 504, 1511–1518, \dodoi{10.1093/mnras/stab929}

\bibitem[{Kimball \& Ivezic(2008)}]{Kimball_2008}
Kimball, A.~E., \& Ivezic, Z. 2008, The Astronomical Journal, 136, 684–712, \dodoi{10.1088/0004-6256/136/2/684}

\bibitem[{{Kimball} {et~al.}(2009){Kimball}, {Knapp}, {Ivezi{\'c}}, {West}, {Bochanski}, {Plotkin}, \& {Gordon}}]{first_sdss}
{Kimball}, A.~E., {Knapp}, G.~R., {Ivezi{\'c}}, {\v{Z}}., {et~al.} 2009, \apj, 701, 535, \dodoi{10.1088/0004-637X/701/1/535}

\bibitem[{Kovari {et~al.}(2023)Kovari, Strassmeier, Kriskovics, Olah, Borkovits, Radvanyi, Granzer, Seli, Vida, \& Weber}]{her}
Kovari, Z., Strassmeier, K.~G., Kriskovics, L., {et~al.} 2023, A star under multiple influences. Magnetic activity in V815 Her, a compact 2+2 hierarchical system.
\newblock \doarXiv{2312.08416}

\bibitem[{Lacy {et~al.}(2020)Lacy, Baum, Chandler, Chatterjee, Clarke, Deustua, English, Farnes, Gaensler, Gugliucci, Hallinan, Kent, Kimball, Law, Lazio, Marvil, Mao, Medlin, Mooley, Murphy, Myers, Osten, Richards, Rosolowsky, Rudnick, Schinzel, Sivakoff, Sjouwerman, Taylor, White, Wrobel, Andernach, Beasley, Berger, Bhatnager, Birkinshaw, Bower, Brandt, Brown, Burke-Spolaor, Butler, Comerford, Demorest, Fu, Giacintucci, Golap, Güth, Hales, Hiriart, Hodge, Horesh, Ivezić, Jarvis, Kamble, Kassim, Liu, Loinard, Lyons, Masters, Mezcua, Moellenbrock, Mroczkowski, Nyland, O’Dea, O’Sullivan, Peters, Radford, Rao, Robnett, Salcido, Shen, Sobotka, Witz, Vaccari, Weeren, Vargas, Williams, \& Yoon}]{vlass}
Lacy, M., Baum, S.~A., Chandler, C.~J., {et~al.} 2020, Publications of the Astronomical Society of the Pacific, 132, 035001, \dodoi{10.1088/1538-3873/ab63eb}

\bibitem[{{Lecavelier Des Etangs} {et~al.}(2011){Lecavelier Des Etangs}, {Sirothia}, {Gopal-Krishna}, \& {Zarka}}]{Etangs+11}
{Lecavelier Des Etangs}, A., {Sirothia}, S.~K., {Gopal-Krishna}, \& {Zarka}, P. 2011, \aap, 533, A50, \dodoi{10.1051/0004-6361/201117330}

\bibitem[{{Leone} {et~al.}(1996){Leone}, {Umana}, \& {Trigilio}}]{MCP}
{Leone}, F., {Umana}, G., \& {Trigilio}, C. 1996, \aap, 310, 271

\bibitem[{McConnell {et~al.}(2020)McConnell, Hale, Lenc, Banfield, Heald, Hotan, Leung, Moss, Murphy, O’Brien, Pritchard, Raja, Sadler, Stewart, Thomson, Whiting, Allison, Amy, Anderson, Ball, Bannister, Bell, Bock, Bolton, Bunton, Chippendale, Collier, Cooray, Cornwell, Diamond, Edwards, Gupta, Hayman, Heywood, Jackson, Koribalski, Lee-Waddell, McClure-Griffiths, Ng, Norris, Phillips, Reynolds, Roxby, Schinckel, Shields, Tremblay, Tzioumis, Voronkov, \& Westmeier}]{racs}
McConnell, D., Hale, C.~L., Lenc, E., {et~al.} 2020, Publications of the Astronomical Society of Australia, 37, \dodoi{10.1017/pasa.2020.41}

\bibitem[{{Morris} \& {Mutel}(1988)}]{rscvn1}
{Morris}, D.~H., \& {Mutel}, R.~L. 1988, \aj, 95, 204, \dodoi{10.1086/114629}

\bibitem[{{Mutel} {et~al.}(1985){Mutel}, {Lestrade}, {Preston}, \& {Phillips}}]{RSCVn}
{Mutel}, R.~L., {Lestrade}, J.~F., {Preston}, R.~A., \& {Phillips}, R.~B. 1985, \apj, 289, 262, \dodoi{10.1086/162886}

\bibitem[{{Narang}(2022)}]{2022MNRAS.515.2015N}
{Narang}, M. 2022, \mnras, 515, 2015, \dodoi{10.1093/mnras/stac1902}

\bibitem[{{Narang} {et~al.}(2021){Narang}, {Manoj}, \& {Ishwara Chandra}}]{2021RNAAS...5..158N}
{Narang}, M., {Manoj}, P., \& {Ishwara Chandra}, C.~H. 2021, Research Notes of the American Astronomical Society, 5, 158, \dodoi{10.3847/2515-5172/ac0fe0}

\bibitem[{{Narang} {et~al.}(2023{\natexlab{a}}){Narang}, {Oza}, {Hakim}, {Manoj}, {Banyal}, \& {Thorngren}}]{exomoon}
{Narang}, M., {Oza}, A.~V., {Hakim}, K., {et~al.} 2023{\natexlab{a}}, \aj, 165, 1, \dodoi{10.3847/1538-3881/ac9eb8}

\bibitem[{Narang {et~al.}(2020)Narang, Manoj, Ishwara Chandra, Lazio, Henning, Tamura, Mathew, Ujwal, \& Mandal}]{Narang_2020}
Narang, M., Manoj, P., Ishwara Chandra, C.~H., {et~al.} 2020, Monthly Notices of the Royal Astronomical Society, 500, 4818–4826, \dodoi{10.1093/mnras/staa3565}

\bibitem[{{Narang} {et~al.}(2023{\natexlab{b}}){Narang}, {Oza}, {Hakim}, {Manoj}, {Tyagi}, {Banerjee}, {Surya}, {Nayak}, {Banyal}, \& {Thorngren}}]{2023MNRAS.522.1662N}
{Narang}, M., {Oza}, A.~V., {Hakim}, K., {et~al.} 2023{\natexlab{b}}, \mnras, 522, 1662, \dodoi{10.1093/mnras/stad1027}

\bibitem[{{Narang} {et~al.}(2024){Narang}, {Puravankara}, {Chandra}, {Banerjee}, {Tyagi}, {Tamura}, {Henning}, {Mathew}, {Lazio}, {Surya}, \& {Nayak}}]{2024MNRAS.529.1161N}
{Narang}, M., {Puravankara}, M., {Chandra}, C.~H.~I., {et~al.} 2024, \mnras, 529, 1161, \dodoi{10.1093/mnras/stae536}

\bibitem[{{Noyola} {et~al.}(2014){Noyola}, {Satyal}, \& {Musielak}}]{Exomoon0}
{Noyola}, J.~P., {Satyal}, S., \& {Musielak}, Z.~E. 2014, \apj, 791, 25, \dodoi{10.1088/0004-637X/791/1/25}

\bibitem[{{Pecaut} \& {Mamajek}(2013)}]{mamajaek}
{Pecaut}, M.~J., \& {Mamajek}, E.~E. 2013, \apjs, 208, 9, \dodoi{10.1088/0067-0049/208/1/9}

\bibitem[{{Pritchard} {et~al.}(2021){Pritchard}, {Murphy}, {Zic}, {Lynch}, {Heald}, {Kaplan}, {Anderson}, {Banfield}, {Hale}, {Hotan}, {Lenc}, {Leung}, {McConnell}, {Moss}, {Raja}, {Stewart}, \& {Whiting}}]{pritchard}
{Pritchard}, J., {Murphy}, T., {Zic}, A., {et~al.} 2021, \mnras, 502, 5438, \dodoi{10.1093/mnras/stab299}

\bibitem[{{Reiners} \& {Basri}(2009)}]{reiners}
{Reiners}, A., \& {Basri}, G. 2009, \aap, 496, 787, \dodoi{10.1051/0004-6361:200811450}

\bibitem[{{Rengelink} {et~al.}(1997){Rengelink}, {Tang}, {de Bruyn}, {Miley}, {Bremer}, {Roettgering}, \& {Bremer}}]{wenss}
{Rengelink}, R.~B., {Tang}, Y., {de Bruyn}, A.~G., {et~al.} 1997, \aaps, 124, 259, \dodoi{10.1051/aas:1997358}

\bibitem[{{Rizzuto} {et~al.}(2017){Rizzuto}, {Mann}, {Vanderburg}, {Kraus}, \& {Covey}}]{k2false}
{Rizzuto}, A.~C., {Mann}, A.~W., {Vanderburg}, A., {Kraus}, A.~L., \& {Covey}, K.~R. 2017, \aj, 154, 224, \dodoi{10.3847/1538-3881/aa9070}

\bibitem[{{Semenko} {et~al.}(2014){Semenko}, {Romanyuk}, {Kudryavtsev}, \& {Yakunin}}]{HD34736}
{Semenko}, E.~A., {Romanyuk}, I.~I., {Kudryavtsev}, D.~O., \& {Yakunin}, I.~A. 2014, Astrophysical Bulletin, 69, 191, \dodoi{10.1134/S1990341314020060}

\bibitem[{{Shimwell} {et~al.}(2017){Shimwell}, {R{\"o}ttgering}, {Best}, {Williams}, {Dijkema}, {de Gasperin}, {Hardcastle}, {Heald}, {Hoang}, {Horneffer}, {Intema}, {Mahony}, {Mandal}, {Mechev}, {Morabito}, {Oonk}, {Rafferty}, {Retana-Montenegro}, {Sabater}, {Tasse}, {van Weeren}, {Br{\"u}ggen}, {Brunetti}, {Chy{\.z}y}, {Conway}, {Haverkorn}, {Jackson}, {Jarvis}, {McKean}, {Miley}, {Morganti}, {White}, {Wise}, {van Bemmel}, {Beck}, {Brienza}, {Bonafede}, {Calistro Rivera}, {Cassano}, {Clarke}, {Cseh}, {Deller}, {Drabent}, {van Driel}, {Engels}, {Falcke}, {Ferrari}, {Fr{\"o}hlich}, {Garrett}, {Harwood}, {Heesen}, {Hoeft}, {Horellou}, {Israel}, {Kapi{\'n}ska}, {Kunert-Bajraszewska}, {McKay}, {Mohan}, {Orr{\'u}}, {Pizzo}, {Prandoni}, {Schwarz}, {Shulevski}, {Sipior}, {Smith}, {Sridhar}, {Steinmetz}, {Stroe}, {Varenius}, {van der Werf}, {Zensus}, \& {Zwart}}]{lotss}
{Shimwell}, T.~W., {R{\"o}ttgering}, H.~J.~A., {Best}, P.~N., {et~al.} 2017, \aap, 598, A104, \dodoi{10.1051/0004-6361/201629313}

\bibitem[{{Shkolnik} {et~al.}(2008){Shkolnik}, {Bohlender}, {Walker}, \& {Collier Cameron}}]{shkolnik+08}
{Shkolnik}, E., {Bohlender}, D.~A., {Walker}, G. A.~H., \& {Collier Cameron}, A. 2008, \apj, 676, 628, \dodoi{10.1086/527351}

\bibitem[{Team {et~al.}(2022)Team, Bean, Bhatnagar, Castro, Meyer, Emonts, Garcia, Garwood, Golap, Villalba, Harris, Hayashi, Hoskins, Hsieh, Jagannathan, Kawasaki, Keimpema, Kettenis, Lopez, Marvil, Masters, McNichols, Mehringer, Miel, Moellenbrock, Montesino, Nakazato, Ott, Petry, Pokorny, Raba, Rau, Schiebel, Schweighart, Sekhar, Shimada, Small, Steeb, Sugimoto, Suoranta, Tsutsumi, van Bemmel, Verkouter, Wells, Xiong, Szomoru, Griffith, Glendenning, \& Kern}]{casa}
Team, T.~C., Bean, B., Bhatnagar, S., {et~al.} 2022, Publications of the Astronomical Society of the Pacific, 134, 114501, \dodoi{10.1088/1538-3873/ac9642}

\bibitem[{Toet {et~al.}(2021)Toet, Vedantham, Callingham, Veken, Shimwell, Zarka, Röttgering, \& Drabent}]{Toet_2021}
Toet, S. E.~B., Vedantham, H.~K., Callingham, J.~R., {et~al.} 2021, Astronomy \&amp; Astrophysics, 654, A21, \dodoi{10.1051/0004-6361/202141163}

\bibitem[{Townsend {et~al.}(2013)Townsend, Rivinius, Rowe, Moffat, Matthews, Bohlender, Neiner, Telting, Guenther, Kallinger, Kuschnig, Rucinski, Sasselov, \& Weiss}]{orionis}
Townsend, R. H.~D., Rivinius, T., Rowe, J.~F., {et~al.} 2013, The Astrophysical Journal, 769, 33, \dodoi{10.1088/0004-637x/769/1/33}

\bibitem[{Trigilio {et~al.}(2023)Trigilio, Biswas, Leto, Umana, Busa, Cavallaro, Das, Chandra, Perez-Torres, Wade, Bordiu, Buemi, Bufano, Ingallinera, Loru, \& Riggi}]{yzceti}
Trigilio, C., Biswas, A., Leto, P., {et~al.} 2023, Star-Planet Interaction at radio wavelengths in YZ Ceti: Inferring planetary magnetic field.
\newblock \doarXiv{2305.00809}

\bibitem[{Vedantham(2020)}]{Vedantham_emission}
Vedantham, H.~K. 2020, Monthly Notices of the Royal Astronomical Society, 500, 3898–3907, \dodoi{10.1093/mnras/staa3373}

\bibitem[{Vedantham {et~al.}(2022)Vedantham, Callingham, Shimwell, Benz, Hajduk, Ray, Tasse, \& Drabent}]{pecRX_hv}
Vedantham, H.~K., Callingham, J.~R., Shimwell, T.~W., {et~al.} 2022, The Astrophysical Journal Letters, 926, L30, \dodoi{10.3847/2041-8213/ac5115}

\bibitem[{{Vedantham} {et~al.}(2020){Vedantham}, {Callingham}, {Shimwell}, {Tasse}, {Pope}, {Bedell}, {Snellen}, {Best}, {Hardcastle}, {Haverkorn}, {Mechev}, {O'Sullivan}, {R{\"o}ttgering}, \& {White}}]{GJ1151}
{Vedantham}, H.~K., {Callingham}, J.~R., {Shimwell}, T.~W., {et~al.} 2020, Nature Astronomy, 4, 577, \dodoi{10.1038/s41550-020-1011-9}

\bibitem[{Walter {et~al.}(2003)Walter, Beck, Morse, \& Wolk}]{tau}
Walter, F.~M., Beck, T.~L., Morse, J.~A., \& Wolk, S.~J. 2003, The Astronomical Journal, 125, 2123–2133, \dodoi{10.1086/368245}

\bibitem[{Wayth {et~al.}(2015)Wayth, Lenc, Bell, Callingham, Dwarakanath, Franzen, For, Gaensler, Hancock, Hindson, Hurley-Walker, Jackson, Johnston-Hollitt, Kapińska, McKinley, Morgan, Offringa, Procopio, Staveley-Smith, Wu, Zheng, Trott, Bernardi, Bowman, Briggs, Cappallo, Corey, Deshpande, Emrich, Goeke, Greenhill, Hazelton, Kaplan, Kasper, Kratzenberg, Lonsdale, Lynch, McWhirter, Mitchell, Morales, Morgan, Oberoi, Ord, Prabu, Rogers, Roshi, Shankar, Srivani, Subrahmanyan, Tingay, Waterson, Webster, Whitney, Williams, \& Williams}]{gleam}
Wayth, R.~B., Lenc, E., Bell, M.~E., {et~al.} 2015, Publications of the Astronomical Society of Australia, 32, \dodoi{10.1017/pasa.2015.26}

\bibitem[{{Webb} {et~al.}(2020){Webb}, {Coriat}, {Traulsen}, {Ballet}, {Motch}, {Carrera}, {Koliopanos}, {Authier}, {de la Calle}, {Ceballos}, {Colomo}, {Chuard}, {Freyberg}, {Garcia}, {Kolehmainen}, {Lamer}, {Lin}, {Maggi}, {Michel}, {Page}, {Page}, {Perea-Calderon}, {Pineau}, {Rodriguez}, {Rosen}, {Santos Lleo}, {Saxton}, {Schwope}, {Tom{\'a}s}, {Watson}, \& {Zakardjian}}]{webbxmm}
{Webb}, N.~A., {Coriat}, M., {Traulsen}, I., {et~al.} 2020, \aap, 641, A136, \dodoi{10.1051/0004-6361/201937353}

\bibitem[{{Webb} {et~al.}(2023){Webb}, {Coriat}, {Traulsen}, {Ballet}, {Motch}, {Carrera}, {Koliopanos}, {Authier}, {de La Calle}, {Ceballos}, {Colomo}, {Chuard}, {Freyberg}, {Garcia}, {Kolehmainen}, {Lamer}, {Lin}, {Maggi}, {Michel}, {Page}, {Page}, {Perea-Calderon}, {Pineau}, {Rodriguez}, {Rosen}, {Santos Lleo}, {Saxton}, {Schwope}, {Tomas}, {Watson}, \& {Zakardjian}}]{xmmdata}
---. 2023, {VizieR Online Data Catalog: XMM-Newton Serendipitous Source Catalogue 4XMM-DR12 (Webb+, 2023)}, VizieR On-line Data Catalog: IX/68. Originally published in: 2020A\&A...641A.136W

\bibitem[{Wenger {et~al.}(2000)Wenger, Ochsenbein, Egret, Dubois, Bonnarel, Borde, Genova, Jasniewicz, Laloë, Lesteven, \& Monier}]{simbad}
Wenger, M., Ochsenbein, F., Egret, D., {et~al.} 2000, Astronomy and Astrophysics Supplement Series, 143, 9–22, \dodoi{10.1051/aas:2000332}

\bibitem[{{Willes} \& {Wu}(2005)}]{white_dwarfs}
{Willes}, A.~J., \& {Wu}, K. 2005, \aap, 432, 1091, \dodoi{10.1051/0004-6361:20040417}

\bibitem[{{Wright} \& {Barlow}(1975)}]{wind}
{Wright}, A.~E., \& {Barlow}, M.~J. 1975, \mnras, 170, 41, \dodoi{10.1093/mnras/170.1.41}

\bibitem[{Yiu {et~al.}(2023)Yiu, Vedantham, Callingham, \& Günther}]{vlass_new_hv}
Yiu, T. W.~H., Vedantham, H.~K., Callingham, J.~R., \& Günther, M.~N. 2023, Radio emission as a stellar activity indicator.
\newblock \doarXiv{2312.07162}

\bibitem[{Zhang {et~al.}(2016)Zhang, Pi, Han, Chang, \& Wang}]{chrom_act}
Zhang, L., Pi, Q., Han, X.~L., Chang, L., \& Wang, D. 2016, Monthly Notices of the Royal Astronomical Society, 459, 854–862, \dodoi{10.1093/mnras/stw668}

\end{thebibliography}
\bibliographystyle{aasjournal}

\end{document}